\documentclass{article}
\usepackage{graphicx}

\begin{document}

\baselineskip=23pt

\begin{center}
\textbf{\Huge Characteristics of profiles of gamma-ray burst pulses
associated with the Doppler effect of fireballs} \vspace{2mm}

\textbf{Yi-Ping Qin$^{1,2,4}$, Zhi-Bin Zhang$^{1,3}$, Fu-Wen Zhang$^{2}$ and
Xiao-Hong Cui$^{1,3}$}

\textbf{$^1$National Astronomical Observatories/Yunnan Observatory, Chinese
Academy of Sciences, P. O. Box 110, Kunming, Yunnan, 650011, P. R. China}

\textbf{$^2$Physics Department, Guangxi University, Nanning, Guangxi,
530004, P. R. China}

\textbf{$^3$ The Graduate School of the Chinese Academy of Sciences}

\textbf{$^4$ E-mail:ypqin@public.km.yn.cn}
\end{center}

\vspace{4mm}

\begin{center}
\textbf{\Large Abstract}
\end{center}

In this paper, we derive in a much detail the formula of count rates, in
terms of the integral of time, of gamma-ray bursts in the framework of
fireballs, where the Doppler effect of the expanding fireball surface is the
key factor to be concerned. Effects arising from the limit of the time delay
due to the limited regions of the emitting areas in the fireball surface and
other factors are investigated. Our analysis shows that the formula of the
count rate of fireballs can be expressed as a function of $\tau $ which is
the observation time scale relative to the dynamical time scale of the
fireball defined by $R_c/c$ where $R_c$ is the fireball radius measured at
an associated local time. The profile of light curves of fireballs depends
only on the relative time scale, entirely independent of the real time scale
and the real size of the objects. It displays in detail how a cutoff tail,
or a turn over, feature (called a cutoff tail problem) in the decay phase of
a light curve can be formed. This feature is a consequence of a hot spot in
the fireball surface, moving towards the observer, and was observed in a few
cases previously. Local pulses suddenly dimming would produce light curves
bearing a certain decaying form (called a standard decaying form) and
exhibiting a sharp feature at their peaks. Light curves arising from
gradually dimming local pulses would be smooth at their peaks, and their
profiles in the decaying phase would obviously deviate from the standard
form when the width of the local pulse is large enough. It is observed that
light curves arising from relatively short local pulses would be the same,
entirely independent of the shape of the latter. Impacts of the rest frame
radiation form and the variance of the form on the profile of light curves
are insignificant, while that on the magnitude of the light curves would be
obvious. By performing fits to the count rate light curves of six sample
sources, we show how to obtain some physical parameters from the observed
profile of the count rate of GRBs and show that there do exist some GRBs
that the profiles of their count rate light curves can be described by the
formula provided. In addition, the analysis reveals that the Doppler effect
of fireballs could lead to a power law relationship between the $FWHM$ of
pulses and energy, which were observed previously by many authors.

\vspace{2mm}

\begin{flushleft}
\end{flushleft}

\vspace{4mm}

\section{Introduction}

Light curves of gamma-ray bursts (GRBs) vary enormously, suggesting that the
temporal activity of the sources would be of a stochastic process (see,
e.g., Fishman et al. 1994). However, some simple bursts with well-separated
structure suggest that they may consist of fundamental units of emission
such as pulses, and some pulses are seen to comprise a fast rise and an
exponential decay (FRED), which can be well represented by an flexible
empirical function (see, e.g., Norris et al. 1996).

Due to the observed great output rate of radiation, GRBs are assumed to
undergo a stage of fireballs which expand relativistically (see, e.g.,
Goodman 1986; Paczynski 1986). As pointed out by Krolik \& Pier (1991),
relativistic bulk motion of the gamma-ray-emitting plasma can account for
some phenomena of GRBs. For example, emission lines would be significantly
broadened due to the curvature of the fireball surface where Doppler
boosting factors varies from point to point (see, e.g., M\'{e}sz\'{a}ros \&
Rees 1998; Heiley et al. 1999; Qin 2003). Promisingly, the observed FRED
structure was found to be interpreted by the curvature effect as the
observed plasma moves relativistically towards us and appears to be locally
isotropic (e.g., Fenimore et al. 1996, hereafter Paper I; Ryde \& Petrosian
2002, hereafter Paper II; Kocevski et al. 2003). Taking into account the
delay of the observational time from the area concerned, estimated by
assuming $\theta \sim 1/\Gamma $, where $\theta $ is the angle to the line
of sight and $\Gamma $ is the Lorentz factor of the expansion, light curves
affected by the curvature effect are available. As derived in detail in
Paper II, a FRED pulse can be well described by the bolometric light curve
of a shell shining continuously, which is
\begin{equation}
F(t)=F_0\int_0^\infty \frac{f(t-x)}{(1+x/t_{ang})^2}dx,
\end{equation}
where $t_{ang}$ is the curvature timescale.

The motivations of the study of this paper are as follows. First, we want to
know how the light curve looks like if only the emission of a hot spot is
concerned, since there could be some on the fireball surface, probably of
the size of $1/\Gamma $, and how the viewing angle of the spot plays a role.
Second, we think it deserved to be investigated that if and how a rest frame
radiation form plays a role on producing the light curve. Third, it would be
interesting to know if and how other factors such as the width and the
structure of local pulses affect the profile of the light curve observed,
which also deserves a detail investigation.

In the following, we will first derive in a much detail the formula suitable
for describing the light curve of fireballs expanding with any velocities so
that it would be applicable to relativistic, sub-relativistic, or even
non-relativistic motions (where, we will pay much of our attention to the
integral limit which might be constrained by the concerned area of the
fireball surface and/or the emission interval of time). Then we will apply
the formula to the case of a local $\delta $ function pulse and show in
detail how the light curve produced by a fraction of the fireball surface
confined by $\theta \leq 1/\Gamma $ differs from that of the whole fireball
surface, and how the light curve would be affected if the patch moving in a
direction other than the line of sight. Later we will study light curves of
different forms and different widths of local pulses. Impacts of the rest
frame radiation form on the profile of light curves will also be
investigated. Applications and discussion will be presented in late sections.

\section{General formula of count rates for expanding fireballs}

There are several papers published studying light curves of relativistically
expanding fireballs (e.g., Paper I; Paper II). We present in the following a
much detailed study on the same issue, where, we do not limit the expanding
speed so that the result would be applicable to relativistic,
sub-relativistic, or even non-relativistic motions, and much of our
attention will be payed to the integral limit which might be constrained by
the concerned area of the fireball surface as well as the emission interval
of time. In this paper we consider the case of a fireball expanding
isotropically with a constant Lorentz factor $\Gamma >1$.

For a radiation independent of direction, the expected flux of a fireball
expanding with a constant Lorentz factor is (Qin 2002, hereafter Paper III)
\begin{equation}
f_\nu (t)=\frac{2\pi }{D^2}\int_{\widetilde{\theta }_{\min }}^{\widetilde{%
\theta }_{\max }}I_\nu (t_\theta ,\nu )R^2(t_\theta )\cos \theta \sin \theta
d\theta ,
\end{equation}
where $\nu $ is the observation frequency; $t$ is the observation time; $D$
is the distance of the fireball to the observer; $\theta $ is the angle, of
the concerned differential surface $ds_\theta $, of the fireball, to the
line of sight; $t_\theta $ is the emission time (in the observer frame),
called local time, of photons which emit from $ds_\theta $; $I_\nu (t_\theta
,\nu )$ is the observer frame intensity; $R(t_\theta )$ is the radius of the
fireball at time $t_\theta $. The integral range of $\theta $, $\widetilde{%
\theta }_{\min }$ and $\widetilde{\theta }_{\max }$, will be determined by
the concerned area of the fireball surface as well as the emission ranges of
the frequency and the local time. Applying the relation between the radius
of the fireball and the observation time (see Paper III) we come to the
following form of the flux:
\begin{equation}
f_\nu (t)=\frac{2\pi c^2[(t-t_c-\frac Dc)\beta +\frac{R_c}c]^2}{D^2}\int_{%
\widetilde{\theta }_{\min }}^{\widetilde{\theta }_{\max }}\frac{I_\nu
(t_\theta ,\nu )\cos \theta \sin \theta }{(1-\beta \cos \theta )^2}d\theta ,
\end{equation}
where $t_c$ and $R_c$ are constants.

Assume that the area of the fireball surface concerned is confined within
\begin{equation}
\theta _{\min }\leq \theta \leq \theta _{\max }
\end{equation}
and the emission time $t_\theta $ is confined within
\begin{equation}
t_c\leq t_{\theta ,\min }\leq t_\theta \leq t_{\theta ,\max },
\end{equation}
and besides them there are no other constraints to the integral limit of (2)
or (3). According to (4) and (5), one can verify that the lower and upper
integral limits of (3) could be determined by
\begin{equation}
\widetilde{\theta }_{\min }=\cos ^{-1}\min \{\cos \theta _{\min },\frac{%
t_{\theta ,\max }-t+\frac Dc}{(t_{\theta ,\max }-t_c)\beta +\frac{R_c}c}\}
\end{equation}
and
\begin{equation}
\widetilde{\theta }_{\max }=\cos ^{-1}\max \{\cos \theta _{\max },\frac{%
t_{\theta ,\min }-t+\frac Dc}{(t_{\theta ,\min }-t_c)\beta +\frac{R_c}c}\},
\end{equation}
respectively (for a detailed derivation, one could refer to Paper III).

As shown by Paper III, $t_\theta $ and $t$ are related by
\begin{equation}
t_\theta =\frac{t-t_c-\frac Dc+\frac{R_c}c\cos \theta }{1-\beta \cos \theta }%
+t_c.
\end{equation}
From (8) one finds that, for any given values of $t$ and $\theta $, $%
t_\theta $ would be uniquely determined. If this value of $t_\theta $ is
within the range of (5), then there will be photons emitted at $t_\theta $
from the small surface area of $\theta $ reaching the observer at $t$ [when $%
\theta $ is within the range of (4), this small area would be included in
the above integral, otherwise it would not]. Obviously, for a certain value
of $\theta $, the range of $t$ depends on the range of $t_\theta $.
Inserting (8) into (5) and applying (4) we obtain
\begin{equation}
\begin{array}{l}
(1-\beta \cos \theta _{\min })t_{\theta ,\min }+(t_c\beta -\frac{R_c}c)\cos
\theta _{\min }+\frac Dc\leq t \\
\leq (1-\beta \cos \theta _{\max })t_{\theta ,\max }+(t_c\beta -\frac{R_c}%
c)\cos \theta _{\max }+\frac Dc
\end{array}
.
\end{equation}
It suggests clearly that observation time $t$ is limited when emission time $%
t_\theta $ is limited.

If during some period the radiation of the fireball is dominated by a
certain mechanism, then within this interval of time the intensity can be
expressed as:
\begin{equation}
I_\nu (t_\theta ,\nu )=I(t_\theta )g_\nu (\nu )=I(t_\theta )\frac{g_{0,\nu
}(\nu _{0,\theta })}{(1-\beta \cos \theta )^3\Gamma ^3},
\end{equation}
where $\nu _{0,\theta }$ is the rest frame emission frequency corresponding
to $\nu $ (they are related by the Doppler effect), $I(t_\theta )$
represents the development of the intensity magnitude in the observer frame,
and $g_\nu (\nu )$ and $g_{0,\nu }(\nu _{0,\theta })$ describe the observer
frame and the rest frame radiation mechanisms, respectively. In deriving the
last equivalency, the Lorentz invariance of $g_\nu (\nu )/\nu ^3$ and the
Doppler effect are applied. Flux (3) then can be written as
\begin{equation}
f_\nu (t)=\frac{2\pi c^2[(t-t_c-\frac Dc)\beta +\frac{R_c}c]^2}{D^2\Gamma ^3}%
\int_{\widetilde{\theta }_{\min }}^{\widetilde{\theta }_{\max }}\frac{%
I(t_\theta )g_{0,\nu }(\nu _{0,\theta })\cos \theta \sin \theta }{(1-\beta
\cos \theta )^5}d\theta ,
\end{equation}
where $\widetilde{\theta }_{\min }$ and $\widetilde{\theta }_{\max }$ are
determined by (6) and (7), respectively, $\nu _{0,\theta }$ and $\nu $ are
related by the Doppler effect, and $t$ is confined by (9).

Light curves of gamma-ray bursts are always presented in terms of count
rates within an energy range. The count rate within energy channel [$\nu _1$%
, $\nu _2$] is determined by
\begin{equation}
\frac{dn(t)}{dt}=\int_{\nu _1}^{\nu _2}\frac{f_\nu (t)}{h\nu }d\nu .
\end{equation}
Applying (11) leads to
\begin{equation}
\frac{dn(t)}{dt}=\frac{2\pi c^2[(t-t_c-\frac Dc)\beta +\frac{R_c}c]^2\int_{%
\widetilde{\theta }_{\min }}^{\widetilde{\theta }_{\max }}\frac{I(t_\theta
)\cos \theta \sin \theta }{(1-\beta \cos \theta )^5}d\theta \int_{\nu
_1}^{\nu _2}\frac{g_{0,\nu }(\nu _{0,\theta })}\nu d\nu }{hD^2\Gamma ^3}.
\end{equation}

Assign
\begin{equation}
\tau _\theta \equiv \frac{t_\theta -t_c}{\frac{R_c}c},
\end{equation}
\begin{equation}
\tau _{\theta ,\min }\equiv \frac{t_{\theta ,\min }-t_c}{\frac{R_c}c},
\end{equation}
\begin{equation}
\tau _{\theta ,\max }\equiv \frac{t_{\theta ,\max }-t_c}{\frac{R_c}c},
\end{equation}
and
\begin{equation}
\tau \equiv \frac{t-t_c-\frac Dc+\frac{R_c}c}{\frac{R_c}c}.
\end{equation}
One would find
\begin{equation}
\tau _\theta =\frac{\tau -(1-\cos \theta )}{1-\beta \cos \theta },
\end{equation}
\begin{equation}
\tau _{\theta ,\min }\leq \tau _\theta \leq \tau _{\theta ,\max },
\end{equation}
and
\begin{equation}
1-\cos \theta _{\min }+(1-\beta \cos \theta _{\min })\tau _{\theta ,\min
}\leq \tau \leq 1-\cos \theta _{\max }+(1-\beta \cos \theta _{\max })\tau
_{\theta ,\max }
\end{equation}
[which is the range of $\tau $ within which the radiation defined within (4)
and (5) is observable].

One could verify that, in terms of the integral of $\tau _\theta $, count
rate (13) becomes
\begin{equation}
C(\tau )=\frac{2\pi R_c^3\int_{\widetilde{\tau }_{\theta ,\min }}^{%
\widetilde{\tau }_{\theta ,\max }}\widetilde{I}(\tau _\theta )(1+\beta \tau
_\theta )^2(1-\tau +\tau _\theta )d\tau _\theta \int_{\nu _1}^{\nu _2}\frac{%
g_{0,\nu }(\nu _{0,\theta })}\nu d\nu }{hcD^2\Gamma ^3(1-\beta )^2(1+k\tau
)^2},
\end{equation}
where $\tau $ is confined by (20),
\begin{equation}
C(\tau )\equiv \frac{dn[t(\tau )]}{d\tau },
\end{equation}
\begin{equation}
\widetilde{I}(\tau _\theta )\equiv I[t_\theta (\tau _\theta )],
\end{equation}
\begin{equation}
k\equiv \frac \beta {1-\beta },
\end{equation}
and $\nu _{0,\theta }$ and $\nu $ are related by
\begin{equation}
\nu _{0,\theta }=\frac{(1+k\tau )(1-\beta )\Gamma }{1+\beta \tau _\theta }%
\nu ,
\end{equation}
while $\widetilde{\tau }_{\theta ,\min }$ and $\widetilde{\tau }_{\theta
,\max }$ are determined by
\begin{equation}
\widetilde{\tau }_{\theta ,\min }=\max \{\tau _{\theta ,\min },\frac{\tau
-1+\cos \theta _{\max }}{1-\beta \cos \theta _{\max }}\}
\end{equation}
and
\begin{equation}
\widetilde{\tau }_{\theta ,\max }=\min \{\tau _{\theta ,\max },\frac{\tau
-1+\cos \theta _{\min }}{1-\beta \cos \theta _{\min }}\},
\end{equation}
respectively.

Taking $t_\theta =t_c$, we then find from (8) that photons emitted from $%
\theta =0$ and at local time $t_c$ would reach the observer at observation
time $t=t_c+D/c-R_c/c$. Thus, the term of $t-t_c-D/c+R_c/c$ in (17)
represents the interval between the observation time of a photon and that of
the photons emitted from $\theta =0$ and at local time $t_c$. Therefore, as
(17) suggests, $\tau $ indicates the above time interval relative to the
dynamical time scale of the fireball defined by $R_c/c$ where $R_c$ is the
fireball radius measured at local time $t_c$ (note that $t_c$ is a local
time based on it $\tau $ is defined). Formula (21) shows that the profile of
count rates of a fireball source is a function of $\tau $. It is independent
of the real time scale $t-t_c-D/c+R_c/c$, and independent of the real size, $%
R_c$, of the source. In other words, no matter how large is the fireball
concerned and how large is the observed timescale concerned, for the profile
of the count rate, only the ratio of the latter to the dynamical time scale
of the fireball plays a role.

\section{Count rate of local $\delta $ function pulses}

Previous studies on the light curve of a local $\delta $ function pulse can
be found in Papers I and II, where, as mentioned above, the limit of the
time delay due to the constraint of the finite emission region is ignored.
Here, we study the same light curve by applying the formula of count rates
derived above, in which, the mentioned constraint is taken into account and,
in addition, other factors possibly ignored by previous studies will be
presented.

Let
\begin{equation}
I(t_\theta )=I_0\delta (t_\theta -t_{\theta ,0})\qquad (t_{\theta ,\min
}\leq t_\theta \leq t_{\theta ,\max }),
\end{equation}
with
\begin{equation}
t_{\theta ,\min }<t_{\theta ,0}<t_{\theta ,\max },
\end{equation}
where $I_0$ is a constant. In terms of $\tau _\theta $, we would get
\begin{equation}
\widetilde{I}(\tau _\theta )=\frac{cI_0}{R_c}\delta (\tau _\theta -\tau
_{\theta ,0})\qquad (\tau _{\theta ,\min }\leq \tau _\theta \leq \tau
_{\theta ,\max })
\end{equation}
and
\begin{equation}
\tau _{\theta ,\min }<\tau _{\theta ,0}<\tau _{\theta ,\max },
\end{equation}
where
\begin{equation}
\tau _{\theta ,0}\equiv \frac{t_{\theta ,0}-t_c}{\frac{R_c}c}.
\end{equation}

One can check that, when
\begin{equation}
1-\cos \theta _{\min }+(1-\beta \cos \theta _{\min })\tau _{\theta ,0}<\tau
<1-\cos \theta _{\max }+(1-\beta \cos \theta _{\max })\tau _{\theta ,0}
\end{equation}
(which is the range of $\tau $ within which the radiation of the local $%
\delta $ function pulse over the concerned area is observable), the
following would be satisfied:
\begin{equation}
\widetilde{\tau }_{\theta ,\min }<\tau _{\theta ,0}<\widetilde{\tau }%
_{\theta ,\max }.
\end{equation}
Inserting (30) into (21) and applying (34) one would get
\begin{equation}
C(\tau )=\frac{2\pi R_c^2I_0\int_{\nu _1}^{\nu _2}\frac{g_{0,\nu }(\nu
_{0,\theta })}\nu d\nu }{hD^2}C_0(\tau ),
\end{equation}
where $\tau $ is confined by (33),
\begin{equation}
C_0(\tau )\equiv \frac{(1+\beta \tau _{\theta ,0})^2(1+\tau _{\theta
,0}-\tau )}{\Gamma ^3(1-\beta )^2(1+k\tau )^2}
\end{equation}
and
\begin{equation}
\nu _{0,\theta }=\frac{(1+k\tau )(1-\beta )\Gamma }{1+\beta \tau _{\theta ,0}%
}\nu .
\end{equation}
It shows that, due to the Doppler effect (or the curvature effect) referring
to the concerned area in the fireball surface, a local $\delta $ function
pulse would produce an observed pulse bearing the shape of $C_0(\tau )$,
where $\tau $ is confined by (33), modified by the rest frame spectrum of
the fireball.

First, let us consider the radiation emitted from the whole fireball surface
(called emitting area 1). In this case we take
\begin{equation}
\theta _{\min }=0\qquad \qquad and\qquad \qquad \theta _{\max }=\frac \pi 2,
\end{equation}
and get from (33) that
\begin{equation}
(1-\beta )\tau _{\theta ,0}<\tau <1+\tau _{\theta ,0}.
\end{equation}
Adopting $\tau _{\theta ,0}=0$, we get $0<\tau <1$. Presented in Fig. 1 are
the curves of $C_0(\tau )$ corresponding to $\Gamma =10$ and $\Gamma =100$,
respectively. The figure shows that, function $C_0(\tau )$ confined by (39)
bears a feature of an exponential decay, and the profile remains the same
for different values of the Lorentz factor. It is interesting that the upper
limit of $\tau $ [see (39)] does not prevent the formation of the
exponential decay tail.

One can verify that, the maximum value of $C_0(\tau )$ is
\begin{equation}
C_{0,p}=\frac 1{\Gamma (\Gamma -\sqrt{\Gamma ^2-1})^2}\frac{R(t_{\theta ,0})%
}{R_c},
\end{equation}
while the width of $C_0(\tau )$ can be determined by
\begin{equation}
\Delta \tau _{FWHM}=\frac{(\Gamma -\sqrt{\Gamma ^2-1})(\sqrt{2\Gamma ^2-1}%
-\Gamma )}{\Gamma ^2-1}\frac{R(t_{\theta ,0})}{R_c},
\end{equation}
and the relation between them is
\begin{equation}
C_{0,p}=\frac{\Gamma ^2-1}{\Gamma (\Gamma -\sqrt{\Gamma ^2-1})^3(\sqrt{%
2\Gamma ^2-1}-\Gamma )}\Delta \tau _{FWHM}.
\end{equation}
(For a detailed derivation one could refer to Appendix A).

Ignoring the effect of the rest frame radiation form (which will be
discussed below), $C_{0,p}$ and $\Delta \tau _{FWHM}$ will serve as the
observed peak and width of the light curve of the local $\delta $ function
pulse. When $\Gamma \gg 1$, one can come to
\begin{equation}
C_{0,p}\simeq 4\Gamma \frac{R(t_{\theta ,0})}{R_c}\qquad (\Gamma \gg 1),
\end{equation}
\begin{equation}
\Delta \tau _{FWHM}\simeq \frac{\sqrt{2}-1}{2\Gamma ^2}\frac{R(t_{\theta ,0})%
}{R_c}\qquad (\Gamma \gg 1),
\end{equation}
and
\begin{equation}
C_{0,p}\simeq \frac{8\Gamma ^3}{\sqrt{2}-1}\Delta \tau _{FWHM}\qquad (\Gamma
\gg 1).
\end{equation}
It shows, both the observed peak and width of the count rate of the local $%
\delta $ function pulse are proportional to the size of the fireball. While
the former rises linearly with the increase of the Lorentz factor, the
latter, as generally known, decays rapidly following the law of $\Gamma
^{-2} $ (see, e.g. Fenimore et al. 1993), which naturally explains why for
many bursts very short time scales, as small as a few $ms$, of pulses have
been observed. For a certain value of the Lorentz factor, quantities $%
C_{0,p} $ and $\Delta \tau _{FWHM}$ are proportional to each other.

Combining (43) and (44) one gets
\begin{equation}
C_{0,p}^2\Delta \tau _{FWHM}\simeq 8(\sqrt{2}-1)(\frac{R(t_{\theta ,0})}{R_c}%
)^3\qquad (\Gamma \gg 1).
\end{equation}
This suggests that, for a same kind of fireball sources, if their difference
is merely due to the Lorentz factor, the product of the square of the peak
count rate and the width of the light curve of their very narrow local
pulses would be the same. This provides a statistical approach to test the
fireball model with pulses of GRBs. For a source, if the intensity of its
pulses remains unchanged, one can expect a high pulse coupling with a small
width while a low pulse coupling with a larger one.

Let $\Delta \tau $ be the interval of the observable time of the local $%
\delta $ function pulse. It would be determined by (39). According to (A6)
and (A8) we get
\begin{equation}
\frac{R(t_{\theta ,0})}{R_c}=\Delta \tau .
\end{equation}
Then (44) becomes
\begin{equation}
\Delta \tau _{FWHM}\simeq \frac{\sqrt{2}-1}{2\Gamma ^2}\Delta \tau \qquad
(\Gamma \gg 1).
\end{equation}
With this one finds that the observed width of the light curve of the local $%
\delta $ function pulse would be several orders of magnitude smaller than
the limit of the observable time when the Lorentz factor is large enough.
This explains why the upper limit of $\tau $ does not prevent the formation
of the exponential decay tail shown in Fig. 1.

Let us turn to study the effect of the limit of the time delay, which refers
to the radiation emitted from a small area with $\theta \leq 1/\Gamma $
(called emitting area 2). Taking
\begin{equation}
\theta _{\min }=0\qquad \qquad and\qquad \qquad \theta _{\max }=\frac
1\Gamma ,
\end{equation}
we get from (33) that
\begin{equation}
(1-\beta )\tau _{\theta ,0}<\tau <1-\cos \frac 1\Gamma +(1-\beta \cos \frac
1\Gamma )\tau _{\theta ,0},
\end{equation}
when $\Gamma $ is large enough which would lead to
\begin{equation}
(1-\beta )\tau _{\theta ,0}<\tau <\frac{1+2\tau _{\theta ,0}}{2\Gamma ^2}%
\qquad (\Gamma \gg 1).
\end{equation}
Adopting $\tau _{\theta ,0}=0$, we get $0<\tau <1/2\Gamma ^2$. The curves of
$C_0(\tau )$ --- $\tau $ for $\Gamma =10$ and $100$ in this case are also
presented in Fig. 1. It shows that, due to the curvature effect, neglecting
the area of $\theta >1/\Gamma $ would lead a light curve, of a local $\delta
$ function pulse, with a cutoff tail in its decay phase, which we call a
cutoff tail problem, suggesting that if only the radiation emitted from the
area of $\theta <1/\Gamma $ is considered, the decay phase of the
corresponding light curve would not be a full exponentially decaying one
(the case of a longer local pulse will be discussed in late sections). [A
similar result can be deduced from the dot lines presented in Fig. 4 of
Paper I.] As shown in Appendix A, the count rate at $\theta =1/\Gamma $ is a
quarter of the peak, and hence the missed part of the light curve is
obviously observable.

What considered above is a patch, with the area of $\theta \leq 1/\Gamma $,
moving towards the observer. What would one expect if the patch moving in a
direction other than the line of sight? Light curves arising from a small
area in the fireball surface, confined by $\theta $ --- $\theta +d\theta $
and $\varphi $ --- $\varphi +d\varphi $, can be calculated by taking $\theta
_{\min }=\theta $ and $\theta _{\max }=\theta +d\theta $ and then
multiplying the resulted $C_0(\tau )$ and $d\varphi /2\pi $. The second step
works due to the highly symmetric nature of the surface. Any forms of
patches can be divided into many small areas confined by $\theta _i$ --- $%
\theta _i+d\theta $ and $\varphi _i$ --- $\varphi _i+d\varphi $ and the
light curves arising from these patches would be obtained by making a sum of
the count rates of the corresponding small areas. Here we consider a simple
case, the radiation emitted from a patch which is confined by
\begin{equation}
\theta _{\min }=\frac 1{2\Gamma },\qquad \qquad \theta _{\max }=\frac
3{2\Gamma };\qquad \qquad \triangle \varphi =\pi ,
\end{equation}
(called emitting area 3). The viewing angle of the center of this patch is $%
\sim 1/\Gamma $ and its size is almost the same as the previous one. We get
from (33) that
\begin{equation}
1-\cos \frac 1{2\Gamma }+(1-\beta \cos \frac 1{2\Gamma })\tau _{\theta
,0}<\tau <1-\cos \frac 3{2\Gamma }+(1-\beta \cos \frac 3{2\Gamma })\tau
_{\theta ,0}.
\end{equation}
Once more, we adopt $\tau _{\theta ,0}=0$. The curves of $C_0(\tau )$ --- $%
\tau $ for $\Gamma =10$ and $100$ in this case are presented in Fig. 1 as
well. Shown in this figure, the feature of cutoff tails is also observed.
Compared with that of the previous patch (emitting area 2), the light curves
last a much longer time, while the amplitudes are much smaller.

Comparing the light curves associated with $\Gamma =10$ and $100$ we find
that, the profiles of the curves corresponding to different Lorentz factors
are not distinguishable, as long as $\Gamma $ is large enough to represent a
relativistic motion.

Compared with that presented in Papers I and II, one finds from (36) that
the factor of $(1+\tau _{\theta ,0}-\tau )$ was previously ignored. When $%
\tau \ll 1+\tau _{\theta ,0}$, this factor is negligible, while when $\tau $
is comparable to $1+\tau _{\theta ,0}$, this factor would play a role.
However, as shown in Fig. 1, a large value of the Lorentz factor will make
the decay phase of the light curve very short, so that the interesting value
of $\tau $ will be very small and then the factor of $(1+\tau _{\theta
,0}-\tau )$ would not be important.

Another factor affecting the light curve is the integral of the rest frame
radiation. When adopting $g_{0,\nu }(\nu _{0,\theta })=\nu _{0,\theta
}^{-\alpha _f}$, one obtains $\int_{\nu _1}^{\nu _2}[g_{0,\nu }(\nu
_{0,\theta })/\nu ]d\nu =g_0(1+k\tau )^{-\alpha _f}$, where $g_0$ is a
constant. A product of this term with the term of $(1+k\tau )^{-2}$ in $%
C_0(\tau )$ is similar to that obtained in Paper I. However, observation
suggests that the common radiation form of GRBs is the so-called Band
function (Band et al. 1993) which was frequently, and rather successfully,
employed to fit the spectra of the sources (see, e.g., Schaefer et al. 1994;
Ford et al. 1995; Preece et al. 1998, 2000). Paper III shows that the
observed radiation form would only be affected slightly by the fireball
Doppler effect. Therefore it is expected that the rest frame radiation of
many fireball sources might bear the Band function form, rather than a power
law one (as adopted in Paper I). In this way, the effect of the radiation
mechanism on the light curve might be different.

Following is the empirical radiation form of GRBs proposed by Band et al.
(1993), the so-called Band function:
\begin{equation}
g_{0,\nu ,B}(\nu _{0,\theta })=\{
\begin{array}{c}
(\frac{\nu _{0,\theta }}{\nu _{0,p}})^{1+\alpha _0}\exp [-(2+\alpha _0)\frac{%
\nu _{0,\theta }}{\nu _{0,p}}]\quad (\frac{\nu _{0,\theta }}{\nu _{0,p}}<%
\frac{\alpha _0-\beta _0}{2+\alpha _0}) \\
(\frac{\alpha _0-\beta _0}{2+\alpha _0})^{\alpha _0-\beta _0}\exp (\beta
_0-\alpha _0)(\frac{\nu _{0,\theta }}{\nu _{0,p}})^{1+\beta _0}\quad (\frac{%
\nu _{0,\theta }}{\nu _{0,p}}\geq \frac{\alpha _0-\beta _0}{2+\alpha _0})
\end{array}
,
\end{equation}
where subscript $B$ represents the word Band, $p$ stands for peak, $\alpha
_0 $\ and $\beta _0$\ are the lower and higher indexes, respectively.
Typical values coming from statistical analysis, of the lower and higher
indexes of the Band function, are $\alpha _0=-1$ and $\beta _0=-2.25$
(Preece et al. 1998, 2000), respectively. As mentioned above, the shape of
rest frame spectra is not significantly changed by the expansion of
fireballs. We take $g_{0,\nu }(\nu _{0,\theta })=g_{0,\nu ,B}(\nu _{0,\theta
})$ and adopt $\alpha _0=-1$ and $\beta _0=-2.25$ to study the effect of the
rest frame spectrum on the light curve of a local $\delta $ function pulse.
Light curves of $C(\tau )$ determined by (35), emitted from the three
emitting areas, calculated within the frequency range of $50\leq \nu /\nu
_{0,p}\leq 100$, for $\Gamma =10$ and $100$, are presented in Fig. 1, where $%
C(\tau )$ is normalized to the peak of the corresponding $C_0(\tau )$. The
same features of cutoff tails are observed. Indicated by the figure, the
impact of the rest frame radiation on the light curve can be obvious (here,
the width of the pulse becomes smaller).

\section{Count rate of general local pulses associated with different
emitting areas in the fireball surface}

In this section, we study the light curve of local pulses with a certain
value of width, associated with different emitting areas in the fireball
surface. A typical and very simple one is a local rectangle pulse, which
will be studied in a much detail. Other forms of local pulses will also be
studied. The limit of the time delay due to the constraint of the finite
emission region of the three emitting areas discussed above will be taken
into account.

\subsection{The case of local rectangle pulses}

To consider a local rectangle pulse we assume
\begin{equation}
I(t_\theta )=\{
\begin{array}{c}
I_0\qquad \qquad \qquad \qquad \qquad (t_{\theta ,\min }\leq t_\theta \leq
t_{\theta ,\max }) \\
0\qquad \qquad (t_\theta <t_{\theta ,\min }\qquad and\qquad t_{\theta ,\max
}<t_\theta )
\end{array}
,
\end{equation}
where $I_0$ is a constant. From (55) we can come to
\begin{equation}
\widetilde{I}(\tau _\theta )=\{
\begin{array}{c}
I_0\qquad \qquad \qquad \qquad \qquad (\tau _{\theta ,\min }\leq \tau
_\theta \leq \tau _{\theta ,\max }) \\
0\qquad \qquad (\tau _\theta <\tau _{\theta ,\min }\qquad and\qquad \tau
_{\theta ,\max }<\tau _\theta )
\end{array}
,
\end{equation}
where, (14), (15), (16) and (23) are applied.

One finds from (26) and (27) that $\tau _{\theta ,\min }\leq \widetilde{\tau
}_{\theta ,\min }$ and $\widetilde{\tau }_{\theta ,\max }\leq \tau _{\theta
,\max }$. With these relations, we come to the following by inserting (56)
into (21):
\begin{equation}
C(\tau )=\frac{2\pi R_c^3I_0\int_{\widetilde{\tau }_{\theta ,\min }}^{%
\widetilde{\tau }_{\theta ,\max }}(1+\beta \tau _\theta )^2(1-\tau +\tau
_\theta )d\tau _\theta \int_{\nu _1}^{\nu _2}\frac{g_{0,\nu }(\nu _{0,\theta
})}\nu d\nu }{hcD^2\Gamma ^3(1-\beta )^2(1+k\tau )^2}.
\end{equation}
This is the formula with which the count rate of a local rectangle pulse can
be calculated.

To focus on how the local width of pulses affects the observed profile, here
we ignore the possible effect from the spectrum and assume a $\delta $
function one:
\begin{equation}
g_{0,\nu }(\nu _{0,\theta })=\delta (\nu _{0,\theta }-\nu _{0,0})\qquad (\nu
_{0,\min }\leq \nu _{0,\theta }\leq \nu _{0,\max }),
\end{equation}
with
\begin{equation}
\nu _{0,\min }<\nu _{0,0}<\nu _{0,\max }.
\end{equation}
Corresponding to $\Gamma >1$, $\beta >0$ would be maintained and will be
applied in the following. Applying (25), we can rewrite (58) as
\begin{equation}
\widetilde{g}_{0,\nu }(\nu ,\tau _\theta ,\tau )=\delta [\frac{(1+k\tau
)(1-\beta )\Gamma }{1+\beta \tau _\theta }(\nu -\nu _0)]\quad (\nu _{\min
}\leq \nu \leq \nu _{\max }),
\end{equation}
where
\begin{equation}
\widetilde{g}_{0,\nu }(\nu ,\tau _\theta ,\tau )\equiv g_{0,\nu }[\nu
_{0,\theta }(\nu ,\tau _\theta ,\tau )],
\end{equation}
\begin{equation}
\nu _0\equiv \frac{1+\beta \tau _\theta }{\Gamma (1-\beta )(1+k\tau )}\nu
_{0,0},
\end{equation}
\begin{equation}
\nu _{\min }\equiv \frac{1+\beta \tau _\theta }{\Gamma (1-\beta )(1+k\tau )}%
\nu _{0,\min }
\end{equation}
and
\begin{equation}
\nu _{\max }\equiv \frac{1+\beta \tau _\theta }{\Gamma (1-\beta )(1+k\tau )}%
\nu _{0,\max }.
\end{equation}
From (59) we find
\begin{equation}
\nu _{\min }<\nu _0<\nu _{\max },
\end{equation}
where, (62), (63) and (64) are applied. According to the property of the $%
\delta $ function, we can rewrite (60) as
\begin{equation}
\widetilde{g}_{0,\nu }(\nu ,\tau _\theta ,\tau )=\frac{1+\beta \tau _\theta
}{(1+k\tau )(1-\beta )\Gamma }\delta (\nu -\nu _0)\quad (\nu _{\min }\leq
\nu \leq \nu _{\max }),
\end{equation}

Suppose
\begin{equation}
\nu _1<\nu _{\min }
\end{equation}
and
\begin{equation}
\nu _{\max }<\nu _2.
\end{equation}
Therefore,
\begin{equation}
\nu _1<\nu _0<\nu _2.
\end{equation}
Replacing $g_{0,\nu }(\nu _{0,\theta })$ in (57) with $\widetilde{g}_{0,\nu
}(\nu ,\tau _\theta ,\tau )$ shown by (66) we obtain
\begin{equation}
C(\tau )=\frac{2\pi R_c^3I_0}{\nu _0hcD^2\Gamma ^4(1-\beta )^3}\frac{\int_{%
\widetilde{\tau }_{\theta ,\min }}^{\widetilde{\tau }_{\theta ,\max
}}(1+\beta \tau _\theta )^3(1-\tau +\tau _\theta )d\tau _\theta }{(1+k\tau
)^3},
\end{equation}
Integrating (70) yields
\begin{equation}
\begin{array}{c}
C(\tau )=\frac{2\pi R_c^3I_0}{5\nu _0hcD^2\Gamma ^4\beta ^2(1-\beta )^3}%
\{[(1+\beta \widetilde{\tau }_{\theta ,\max })^5-(1+\beta \widetilde{\tau }%
_{\theta ,\min })^5](1+k\tau )^{-3} \\
-\frac 54(1-\beta )[(1+\beta \widetilde{\tau }_{\theta ,\max })^4-(1+\beta
\widetilde{\tau }_{\theta ,\min })^4](1+k\tau )^{-2}\},
\end{array}
\end{equation}
where $\beta >0$ is applied, and $\widetilde{\tau }_{\theta ,\min }$ and $%
\widetilde{\tau }_{\theta ,\max }$ are determined by (26) and (27),
respectively. This is the formula for calculating the count rate of the
local rectangle pulse which ignores the effect of spectra.

To make the plot of $C(\tau )$, we consider radiation from the three
emitting areas discussed in last section: emitting areas 1, 2 and 3, which
correspond to the whole fireball surface, the small area of $\theta \leq
1/\Gamma $ and that confined by (52), respectively. In the first case, (38)
is applicable, and then from (26), (27) and (20) we get
\begin{equation}
\widetilde{\tau }_{\theta ,\min }=\max \{\tau _{\theta ,\min },\tau -1\},
\end{equation}
\begin{equation}
\widetilde{\tau }_{\theta ,\max }=\min \{\tau _{\theta ,\max },\frac \tau
{1-\beta }\},
\end{equation}
and
\begin{equation}
(1-\beta )\tau _{\theta ,\min }\leq \tau \leq 1+\tau _{\theta ,\max },
\end{equation}
respectively. In the second case, one should apply (49) and then gets
\begin{equation}
\widetilde{\tau }_{\theta ,\min }=\max \{\tau _{\theta ,\min },\frac{\tau
-1+\cos \frac 1\Gamma }{1-\beta \cos \frac 1\Gamma }\},
\end{equation}
\begin{equation}
\widetilde{\tau }_{\theta ,\max }=\min \{\tau _{\theta ,\max },\frac \tau
{1-\beta }\},
\end{equation}
and
\begin{equation}
(1-\beta )\tau _{\theta ,\min }\leq \tau \leq 1-\cos \frac 1\Gamma +(1-\beta
\cos \frac 1\Gamma )\tau _{\theta ,\max }.
\end{equation}
In the third case, (52) would be applied, and then we obtain
\begin{equation}
\widetilde{\tau }_{\theta ,\min }=\max \{\tau _{\theta ,\min },\frac{\tau
-1+\cos \frac 3{2\Gamma }}{1-\beta \cos \frac 3{2\Gamma }}\}
\end{equation}
\begin{equation}
\widetilde{\tau }_{\theta ,\max }=\min \{\tau _{\theta ,\max },\frac{\tau
-1+\cos \frac 1{2\Gamma }}{1-\beta \cos \frac 1{2\Gamma }}\},
\end{equation}
and
\begin{equation}
1-\cos \frac 1{2\Gamma }+(1-\beta \cos \frac 1{2\Gamma })\tau _{\theta ,\min
}\leq \tau \leq 1-\cos \frac 3{2\Gamma }+(1-\beta \cos \frac 3{2\Gamma
})\tau _{\theta ,\max }.
\end{equation}

We employ the same Band function form of radiation with $\alpha _0=-1$ and $%
\beta _0=-2.25$, as adopted above, to illustrate profiles of $C(\tau )$
determined by (57), where $g_{0,\nu }(\nu _{0,\theta })$ would be replaced
by $g_{0,\nu ,B}(\nu _{0,\theta })$ shown in (54) and $\nu _{0,\theta }$ is
related with $\nu $ by (25). Light curves arising from the three emitting
areas, calculated within the frequency range of $50\leq \nu /\nu _{0,p}\leq
100$, are presented in Fig. 2, where we take $2\pi R_c^3I_0/hcD^2=1$, $%
\Gamma =10$ and $\tau _{\theta ,\min }=0$, and adopt $\tau _{\theta ,\max
}=0.2$, $2$ and $20$, respectively. The figure shows explicitly a structure
of FRED for the light curve arising from emitting area 1, suggesting that,
such pulses can arise from a fireball surface when the local pulse involved
lasts an interval of time (in contrast with it, there exists only an
exponential decay phase in the light curve of a local $\delta $ function
pulse, for which no rising phase can be seen). One finds that while the
decay phase is due to the curvature effect, the rising portion of FRED
pulses is produced by the width of the local pulse, which was already known
(see, e.g., Paper I). (For a detailed analysis of the rising phase one could
refer to Appendix B). One finds that the less the width of the local pulse,
the narrower the observed rising phase. We suspect that, for many GRBs, FRED
pulses observed might mainly be due to the expanding motion of fireballs.
When taking different values of $\Gamma $, we find almost the same form of
curves, suggesting that the character of FRED is a consequence of the
expanding motion of fireballs as long as the motion is relativistic, no
matter how large the Lorentz factor is. The cutoff tail feature is also
observed in this figure (in the light curve arising from emitting area 2),
with the longer the local pulse the less obvious the feature. When the local
pulse is long enough, the cutoff tail feature would no more be visible, but
instead, light curves arising from emitting area 2 in the decay phase would
drop more rapidly than those from emitting area 1. In addition, for the
patch moving along a direction with a small angle to the line of sight, the
feature of the cutoff tail would not be visible when the local pulse is not
very short (for extremely short local pulses, their light curves would
approach to those of Fig. 1), and, interesting enough, the interval between
the start and peak of its count rate is much larger than that associated
with other two emitting areas. Plotting the light curves of (71), where the
rest frame radiation form is ignored, we find almost the same result (the
plot is omitted), suggesting that, to produce the profile of the light
curves of the cases considered here, the rest frame radiation form does not
play an important role (a more detailed discussion on the influence of the
rest frame radiation form on the light curve, both for the magnitude and the
profile, will be presented below).

\subsection{The case of other forms of local pulses}

Here we study if different forms of local pulses would lead to much
different forms of expected light curves. In the following we consider
several forms of local pulses other than the rectangle one.

First, let us consider a local pulse with an exponential rise and an
exponential decay, which is written as
\begin{equation}
\widetilde{I}(\tau _\theta )=I_0\{
\begin{array}{c}
\exp (\frac{\tau _\theta -\tau _{\theta ,0}}\sigma )\qquad \qquad \qquad
(\tau _{\theta ,\min }\leq \tau _\theta \leq \tau _{\theta ,0}) \\
\exp (-\frac{\tau _\theta -\tau _{\theta ,0}}\sigma )\qquad \qquad \qquad
\qquad \qquad (\tau _{\theta ,0}<\tau _\theta )
\end{array}
.
\end{equation}
This form of intensity belongs to the class of gradually shining and
gradually dimming local pulses. [One could observe that, when $\left| \tau
_\theta -\tau _{\theta ,0}\right| /\sigma \ll 1$, intensity (81) would
approach to that of linear functions.] Note that, since $1+\beta \tau
_\theta =R(t_\theta )/R_c>0$ (see Appendix A), $\tau _\theta >-1/\beta $.
That provides a constraint to the lower limit of $\tau _\theta $, i. e., $%
\tau _{\theta ,\min }>-1/\beta $.

We employ the same Band function radiation form with $\alpha _0=-1$ and $%
\beta _0=-2.25$ to make the light curve. The count rate is determined by
(21), where $g_{0,\nu }(\nu _{0,\theta })$ would be replaced by $g_{0,\nu
,B}(\nu _{0,\theta })$ [see (54)], and $\nu _{0,\theta }$ is related with $%
\nu $ by (25). We also consider the three cases associated with emitting
areas 1, 2, and 3, respectively. For emitting area 1, $\widetilde{\tau }%
_{\theta ,\min }$ and $\widetilde{\tau }_{\theta ,\max }$ would be
determined by (72) and (73), respectively, while $\tau $ would be confined
by (74); for emitting area 2, $\widetilde{\tau }_{\theta ,\min }$ and $%
\widetilde{\tau }_{\theta ,\max }$ would be calculated with (75) and (76),
respectively, while $\tau $ would be confined by (77); for emitting area 3, $%
\widetilde{\tau }_{\theta ,\min }$ and $\widetilde{\tau }_{\theta ,\max }$
would be determined by (78) and (79), respectively, while the range of $\tau
$ would be within that of (80). As the local emission time is limited by $%
\tau _\theta >-1/\beta $, it is impossible to take a negative infinity value
of $\tau _{\theta ,\min }$ and therefore the interval between $\tau _{\theta
,0}$ and $\tau _{\theta ,\min }$ must be limited. Here we assign $\tau
_{\theta ,0}=10\sigma +\tau _{\theta ,\min }$ so that the interval between $%
\tau _{\theta ,0}$ and $\tau _{\theta ,\min }$ would be large enough to make
the rising part of the local pulse close to that of the exponential pulse.

Profiles of $C(\tau )$ determined by (21), of this local pulse, calculated
within the frequency range of $50\leq \nu /\nu _{0,p}\leq 100$, for the
radiation emitted from the three emitting areas are also presented in Fig.
2, where we take $\Gamma =10$, $\tau _{\theta ,\min }=0$, $2\pi
R_c^3I_0/hcD^2=1$, and adopt $\sigma =0.2$, $2$ and $20$, respectively.

Shown in the figure, the structure of FRED is also observed, suggesting that
the character of FRED is independent of the local structure of pulses. It is
indeed a consequence of the expanding motion of fireballs. Different from
those of the local rectangle pulses, the curves presented here are less
sharp around the position of the peak count rate. A turn over, instead of a
cutoff, tail is observed in the light curves associated with this kind of
local pulses, and the shorter the local pulse, the more obvious the turn
over feature. For longer local pulses, the corresponding profiles seem to be
the same (a direct comparison will be made below).

In addition, we consider three other local pulses. The first is a local
pulse with an exponential rise
\begin{equation}
\widetilde{I}(\tau _\theta )=I_0\exp (\frac{\tau _\theta -\tau _{\theta
,\max }}\sigma )\qquad \qquad \qquad (\tau _{\theta ,\min }\leq \tau _\theta
\leq \tau _{\theta ,\max }),
\end{equation}
and the second is a local pulse with an exponential decay
\begin{equation}
\widetilde{I}(\tau _\theta )=I_0\exp (-\frac{\tau _\theta -\tau _{\theta
,\min }}\sigma )\qquad \qquad \qquad \qquad \qquad (\tau _{\theta ,\min
}\leq \tau _\theta ),
\end{equation}
and the third is a local pulse with a Gaussian form
\begin{equation}
\widetilde{I}(\tau _\theta )=I_0\exp [-(\frac{\tau _\theta -\tau _{\theta ,0}%
}\sigma )^2]\qquad \qquad \qquad \qquad \qquad (\tau _{\theta ,\min }\leq
\tau _\theta ).
\end{equation}
Among them, the first is a gradually shining and suddenly dimming local
pulse, and the second is a suddenly shining and gradually dimming local
pulse, while the third is a gradually shining and relatively fast dimming
local pulse when $\tau _{\theta ,\min }<\tau _{\theta ,0}$.

We take the same parameters as those adopted in the case of local pulse (81)
to study the light curves associated with these three local pulses, where
for the first and third local pulses we assign $\tau _{\theta ,\max
}=10\sigma +\tau _{\theta ,\min }$ and $\tau _{\theta ,0}=10\sigma +\tau
_{\theta ,\min }$ respectively. The curves of $C(\tau )$ arising from the
three emitting areas associated with these three local pulses are presented
in Fig. 2 as well. We find that, light curves associated with local pulse
(84) are similar to those associated with local pulse (81); in the case of
local pulse (83), the cutoff or the turn over feature is visible only when
the local pulse is short enough, otherwise it would no longer appear; for
local pulse (82), light curves are even sharper than those associated with
the rectangle local pulse, where the turn over feature is also observed, and
is more obvious than that shown in the case of local pulse (81).

One can conclude from the figure that sudden dimming local pulses (either
short or long) would give rise to sharp features of the light curves (see
panels of the first and third rows in Fig. 2); gradually dimming local
pulses would give birth to smooth light curves (see panels of the second and
fourth rows in Fig. 2); relatively fast dimming local pulses would produce
less smooth (or less sharp) light curves (see panels of the fifth row in
Fig. 2). For relatively short light curves, the cutoff or the turn over tail
feature would be obvious (see panels of the first column in Fig. 2).

It is noticed that, even though sudden dimming local pulses would give rise
to sharp features of the light curves, suddenly shining local pulses would
not. This must be due to the fact that the former would give up their roles
to the curvature effect after the dimming begins, while the latter would not.

\section{Impact of other factors on the profile of light curves}

In this section, we will study impacts of other possible factors (other than
different emitting areas discussed above) on the expected light curve of
fireballs. How the width and structure of local pulses as well as the rest
frame radiation form would affect the profile of light curves will be
investigated. To focus on these effects, we consider here and in the late
sections only the radiation emitted from the whole fireball surface (the
so-called emitting area 1), for which, (38) would be applied.

\subsection{Influence of the width of local pulses}

The influence of the width of local pulses on the profile of light curves
can be shown by plotting in a same figure various light curves corresponding
to different widthes of local pulses. In doing this, light curves should be
normalized both in the magnitude and time scale so that they are visually
comparable. In the following, for each curve, the magnitude of count rates
would be normalized to a unit and the relative time, the variable, $\tau $,
would be re-scaled so that the peak count rate is located at $\tau =0$ and
the $FWHM$ of the decay portion is located at $\tau =0.2$.

To illustrate this effect in a more general manner, besides those local
pulses discussed in last section, three other forms of local pulses are
considered. The first is a local pulse with a power law rise and a power law
decay which is assumed to be
\begin{equation}
\widetilde{I}(\tau _\theta )=I_0\{
\begin{array}{c}
(\frac{\tau _\theta -\tau _{\theta ,\min }}{\tau _{\theta ,0}-\tau _{\theta
,\min }})^\mu \qquad \qquad \qquad \qquad \qquad (\tau _{\theta ,\min }\leq
\tau _\theta \leq \tau _{\theta ,0}) \\
(1-\frac{\tau _\theta -\tau _{\theta ,0}}{\tau _{\theta ,\max }-\tau
_{\theta ,0}})^\mu \qquad \qquad \qquad \qquad (\tau _{\theta ,0}<\tau
_\theta \leq \tau _{\theta ,\max })
\end{array}
.
\end{equation}
[When $\mu =1$, local pulse (85) would become a linear rise and a linear
decay one.] For this local pulse we find $\tau _{\theta ,FWHM1}=2^{-1/\mu
}\tau _{\theta ,0}+(1-2^{-1/\mu })\tau _{\theta ,\min }$ and $\tau _{\theta
,FWHM2}=2^{-1/\mu }\tau _{\theta ,0}+(1-2^{-1/\mu })\tau _{\theta ,\max }$.
In this section, we consider only the case of $\mu =2$. Thus, the $FWHM$ of
this local pulse would be $\Delta \tau _{\theta ,FWHM}=(1-1/\sqrt{2})(\tau
_{\theta ,\max }-\tau _{\theta ,\min })$, which leads to $\tau _{\theta
,\max }=\Delta \tau _{\theta ,FWHM}/(1-1/\sqrt{2})+\tau _{\theta ,\min }$.
The second is a local pulse with a power law rise following
\begin{equation}
\widetilde{I}(\tau _\theta )=I_0(\frac{\tau _\theta -\tau _{\theta ,\min }}{%
\tau _{\theta ,\max }-\tau _{\theta ,\min }})^\mu \qquad \qquad \qquad
\qquad \qquad (\tau _{\theta ,\min }\leq \tau _\theta \leq \tau _{\theta
,\max }),
\end{equation}
and the third is that with a power law decay which is written as
\begin{equation}
\widetilde{I}(\tau _\theta )=I_0(1-\frac{\tau _\theta -\tau _{\theta ,\min }%
}{\tau _{\theta ,\max }-\tau _{\theta ,\min }})^\mu \qquad \qquad \qquad
\qquad (\tau _{\theta ,\min }<\tau _\theta \leq \tau _{\theta ,\max }).
\end{equation}
In the case of $\mu =2$, the relation of $\tau _{\theta ,\max }=\Delta \tau
_{\theta ,FWHM}/(1-1/\sqrt{2})+\tau _{\theta ,\min }$ holds for the two
latter local pulses. We observe that, the first belongs to the class of
gradually shining and gradually dimming local pulses, the second is a
gradually shining and suddenly dimming local pulse, and the third is a
suddenly shining and gradually dimming local pulse.

We employ the same Band function radiation form with $\alpha _0=-1$ and $%
\beta _0=-2.25$ to make the light curve. The count rate is determined by
(21), where $g_{0,\nu }(\nu _{0,\theta })$ would be replaced by $g_{0,\nu
,B}(\nu _{0,\theta })$ [see (54)], and $\nu _{0,\theta }$ is related with $%
\nu $ by (25). We take $\Gamma =10$ and $\tau _{\theta ,\min }=0$, and adopt
$\Delta \tau _{\theta ,FWHM}=0.02$, $0.2$, $2$ and $20$, respectively, to
make the profiles of the light curves of these local pulses, calculated
within the frequency range of $50\leq \nu /\nu _{0,p}\leq 100$ (see Fig. 3).
For local pulse (85), $\tau _{\theta ,0}=\tau _{\theta ,\max }/2$ is adopted.

Shown in Fig. 3 are the normalized and re-scaled curves associated with
various widthes of local pulses for the five intensities studied in last
section (where, except the magnitude and the time scale and the width of
local pulses, all parameters are the same as those adopted in calculating
the corresponding curves in Fig. 2) as well as the three intensities
presented in this section. We find in Fig. 3 that, for suddenly dimming
local pulses (see panels of the first column in the first, second and fourth
rows), the shape, a concave curve, of the decay phase keeps to be the same
for various values of the local pulse width (we therefore call it the
standard decay curve); for a relatively fast dimming local pulse such as the
Gaussian local pulse, the shape of the decay phase light curve slightly
deviates from the standard form and keeps almost unchanged (see panel of the
first column third row); for gradually dimming local pulses, the profile of
the light curve in the decay phase varies with the local pulse width; for
narrow local pulses, the width of the rising portion of the corresponding
light curve, relative to that of the decay phase, is sensitive to the local
pulse width, where the smaller the local pulse width the narrower the rising
part of the light curve; for longer local pulses the relative width
(relative to that of the decay part) of the rising portion of the light
curve would no longer depend on the local pulse width, but instead, would
keep to be unchanged. It is interesting that both convex and concave curves
in the rising portion of the light curve could be observed in Fig. 3, which
depend on the shape of local pulses.

To show in a much detail how the relative width of the rising portion,
relative to that of the decay phase, of light curves is affected by the
local pulse width, we present Fig. 4, where, $FWHM1$ is the $FWHM$ of the
rising portion, and $FWHM2$ is that of the decay phase. We find in this
figure that the relative width, $FWHM1/FWHM2$, is sensitive to the local
width as long as the latter is small enough. When the latter is sufficiently
large, e.g. $\sigma =1$, the former would remain unchanged, and in this
situation, the two widthes, $FWHM1$ and $FWHM2$, would be proportional to
each other. This conclusion holds for any forms of local pulses. The upper
limit of the sensitivity of the relative width to the local width differs
for various forms of local pulses. For all kinds of local pulses, the value
of $FWHM1/FWHM2$ would never exceed 1.3, which might be a criterion to check
if a pulse arises from the emission of the whole fireball surface. Within
the sensitivity range, the relative width would be uniquely related with the
local width for any forms of the local pulse. In this situation, the former
would be able to serve as an indicator of the latter.

Listed in Table 1 are the values of $FWHM1/FWHM2$ of the light curves of the
eight local intensities analyzed in Fig. 4, for some typical values of the
local width.

Note that, the words of ``small width'' and ``large width'' mentioned here
are in terms of the relative time scale $\tau $. As explained in section 2,
even for a very thick shell which might produce a large time scale of a
local pulse, if the size of the source is sufficiently large (such as the
afterglow of GRBs) so that the dynamical time scale of the fireball which is
defined with the fireball radius is large enough, when the ratio of the
former time scale to the latter time scale is very small, the pulse would
still be regarded as a short one and the ratio of the width of the rising
portion of the light curve to that of the decay phase would be small, and,
in this situation, the form of the local pulse employed to fit the light
curve would not be important (see what discussed below). In particular, the
profile of short pulses presented in Fig. 3 (i.e., the solid lines there)
could be observed at very late epoches if the fireball model is applicable
to a source and if the time scale of shocks is very small compared with the
dynamical time scale of the fireball. In reverse, a shock creating very
short time scale of local pulses could also lead to the profile of long
pulses shown in Fig. 3 (e.g., the dot lines there) if the dynamical time
scale of the fireball is small enough (such as in the period of the trigger
time of bursts). A conclusion of these is that profiles of the curves in
Fig. 3 could be observed in any periods of the light curves of GRBs if these
sources can be described by the fireball model.

\subsection{Influence of the shape of local pulses}

To show how the shape of local pulses plays a role in producing the expected
light curves of fireball sources we present Fig. 5. Displayed in this figure
are the same curves of the panels of the first two rows of Fig. 3 (those of
the panels of the last two rows are omitted due to the similarity), where
light curves of different kinds, arising from a same local pulse width, are
plotted in the same panel. We find that, the smaller the width of local
pulses, the more similar the profile of light curves of various kinds of
local pulses. When the local pulse is short enough (say $\sigma =0.02$ or
smaller), light curves arising from different forms of local pulses would
not be distinguishable, for which, the shape of the light curve in the decay
portion would be the same as those arising from suddenly dimming local
pulses (the standard decaying form; see panels of the first column first and
second rows of Fig. 3). This enables us to fit a light curve with a very
short width of its rising portion, relative to that in the decay part, with
any forms of local pulses, such as a local rectangle pulse, without causing
a significant difference (in other words, one can fit such light curves
quite satisfactorily without knowing the real form of the local pulse). This
becomes one of the conclusions of this paper. Panels of the third and fourth
columns of Fig. 5 show that, when the local pulse width is large enough, a
certain kind of local pulses would produce a definite form of light curves
and the profile of the curves would remain the same for different values of
the local pulse width.

It is noticed that, the standard decaying curve is just the same as that
produced by a very short local pulse, and for the latter, the decay phase
must merely be due to the geometry of the fireball surface. Thus, the
standard shape is associated with nothing but the pure curvature effect.

Since the decay phase of suddenly dimming local pulses (see solid lines in
panels of the first and third rows of Fig. 5) bears the standard shape, the
figure shows obviously that the decay curve of gradually dimming local
pulses betrays the standard form in the manner that it is convex before $%
FWHM2$ and is concave after $FWHM2$ (see, e.g., solid lines in panels of the
second, third and fourth columns of the second and fourth rows of Fig. 5).
This manner will hold as long as the local pulse width is not very small
(say, $\sigma =0.2$ or larger).

\subsection{Influence of the rest frame radiation form}

Let us turn to study the impact of the rest frame radiation form on the
expected light curve of fireballs. As a general radiation form observed, the
Band function, for which, some sets of typical values of the indexes would
represent certain mechanisms (see Band et al. 1993), will be employed in the
following analysis. We will first investigate if different indexes would
lead to a much different profile of light curves, and then will study how
the evolution of the indexes is at work in producing the light curve.

The impact of indexes will be shown when light curves arising from the local
pulses defined by (81) and (82) are plotted by taking different values of
the indexes of the Band function. The previously adopted one, the Band
function with $\alpha _0=-1$ and $\beta _0=-2.25$, will be compared with
that of $\alpha _0=-0.5$ and $\beta _0=-3$ and that of $\alpha _0=-1.5$ and $%
\beta _0=-2$. Displayed in Fig. 6 are these curves. We find that, while the
shapes of the light curves seem quite similar for different values of the
indexes, their magnitudes differ obviously.

To tell how the shapes of the light curves are affected, we once more plot
these curves in the manner adopted in plotting Fig. 3, where these light
curves are normalized and their variables, $\tau $, are re-scaled so that
the peak count rate is located at $\tau =0$ and the $FWHM$ of the decay
portion is located at $\tau =0.2$. Presented in Fig. 7 are these normalized
and re-scaled curves. It shows that, for relatively short local pulses, the
profile of light curves would well keep its shape; for relatively longer
local pulses, the profile would be mildly affected and the difference would
be hardly detectable. We come to the conclusion that the profile is not
significantly affected by the rest frame radiation form.

Another factor possibly affecting the profile of the light curve is the
evolution of $\alpha _0$, which was often observed. Let us consider an
evolution of the index ranging from $\alpha _0=-0.5$ to $\alpha _0=-1.5$. We
once more study light curves arising from the local pulses defined by (81)
and (82). For intensity (81), we assume $\alpha _0=-1.5+\exp [-(\tau _\theta
-\tau _{\theta ,\min })]$, and for intensity (82), we assume $\alpha
_0=-0.5\exp [\ln 3\times (\tau _\theta -\tau _{\theta ,\min })/(\tau
_{\theta ,\max }-\tau _{\theta ,\min })]$. In this way, for both cases one
would get $\alpha _0=-0.5$ at the beginning of the local pulse and get $%
\alpha _0=-1.5$ at the end. We calculate the count rate with these
relations, where, other parameters are the same as those adopted in making
Figs. 6 and 7. Similar results are obtained. The magnitude of the light
curve is affected by the evolution of the index as well, as shown by those
curves in Fig. 6. However, when plotting the normalized and re-scaled light
curve as done in Fig. 7, we come to the same conclusion. (The figures are
omitted due to the similarity to Figs. 6 and 7.)

\section{Application to some GRBs}

According to the above analysis, we are aware that, if a gamma-ray burst is
under the stage of fireballs, the profile of its light curve would be, or
would be similar to, one of the curves of Fig. 3. If the width of the rising
portion relative to that of the decay phase is very small, then the profile
would be well fitted by one of the curves of the panels of the first column
of Fig. 5, or by one close to those. In this situation, if the type of the
corresponding local pulse is identified, then the local pulse width obtained
by fit would be well determined since it is sensitive to the relative width
observed for any types concerned (see Fig. 4). If the relative width is
large enough, then the type (suddenly dimming or gradually dimming) of the
corresponding local pulse would be well distinguished (see Fig. 5), although
the local pulse width would no longer be determined (see Fig. 4).

Here we study the profile of light curves of several GRBs (GRB 910721, GRB
920925, GRB 930612, GRB 941026, GRB 951019, and GRB 951102B), which light
curves are likely to be those of FRED pulses, trying to find out if the
light curves could be represented by any of the curves discussed above, and
if so, find out what could we obtain from the analysis.

Count rates of these sources are available in the web site of BATSE, where
the presented counts are within the bin of $64ms$ for four energy channels
(channel 1, $25-55kev$; channel 2, $55-110kev$; channel 3, $110-320kev$;
channel 4, $>320kev$). It has been already known that pulses of GRBs show a
tendency to self-similarity across energy bands (see, e.g., Norris et al.
1996). Thus, we would study the count rate of only one of the channels. The
one selected is channel 3, as the break energy of most GRBs could be found
within this range (see Preece et al. 2000), and therefore count rates of
this channel would be large enough for a statistical study. For each source,
we assume its signal data covers the range of $t_{\min }\leq t\leq t_{\max }$%
, where $t_{\max }-t_{\min }=2T_{90}$, and $t_{\min }$ is at $T_{90}/2$
previous to the start of $T_{90}$. Data beyond this range, called sample 1,
would be taken to find the fit of the background. These background data
would first be smoothed with the DB3 wavelet (the first-class decomposition)
with the MATLAB software, called sample 2, and then would be fitted with a
linear function. This background fit would be applied to the signal interval
and would be taken as the background count rate there.

Data within the signal interval, called sample 3, subtracting the background
counts would be taken as the signal data, called sample 4. First, sample 4
would be smoothed with the DB3 wavelet in the level of the third-class
decomposition, and with these smoothed data, called sample 5, we would get
primary values of the magnitude and position of the peak count rate, and
then with these peak count rate parameters we would find the corresponding
position of the $FWHM$ in the decay phase. Second, sample 4 would be
smoothed with the DB3 wavelet in the first-class decomposition level, and
these smoothed data, called sample 6, would be normalized to the peak count
rate and re-scaled to the positions of the peak and the $FWHM$ (the former
would be assigned to be $0$ and the latter $0.2$), called sample 7. We will
compare the data of sample 7 with several theoretical curves discussed
above, and among them the one that is the closest to the data would be
selected. We will perform a fit to the data of sample 6 with the selected
curve, where, the least square method would be used. When performing the
fit, not only parameters of the curve, but also the magnitude as well as the
time scale and the origin of time for the curve would be free. With the
fitting curve, we would obtain the final values of the magnitude and
position of the peak count rate and the $FWHM$ in the decay phase, for
sample 6. With these peak count rate and $FWHM$ parameters, sample 6 would
once more be normalized and re-scaled in the same way performed above, which
is called sample 8. Data of sample 8 and the fitting curve will be presented
in a same figure to show the result of the fit. The goodness of fit would be
described by the statistics $\chi ^2$ which is defined by $\chi ^2\equiv
\sum_{i=1}^n(C_{ob,i}-C_i)^2/C_i$, where $C_{ob,i}$ and $C_i$ are the
observed and expected counts, respectively, within the $i$the bin, and $n$
is the total number of bins. It is noticed that, in terms of statistics, the
fluctuation of $C_{ob,i}$ must be due to both the signal and background
counts. Sample 6 itself is not suitable to calculate $\chi ^2$ defined
above. Therefore, in calculating the statistics, data of sample 6 plus the
background fit would be employed to determine $C_{ob,i}$, and the fitting
curve plus the same background fit would be employed to determine $C_i$
(note that, in doing so, a zero value of $C_i$ will no longer appears).

Let us study the count rates of GRB 930612 (\#2387) in detail. The duration
of GRB 930612 is $T_{90}=41.984s$, and the start time of its $T_{90}$ is $%
2.112s$. One finds $t_{\min }=-18.88s$ and $t_{\max }=65.088s$.

First, let us check if there is a self-similarity across energy bands for
this burst. This would be done when the profiles of the light curves of
different channels are plotted in a same figure and are compared. For this
source, a FRED pulse light curve is visible in channels 1, 2, and 3 (in
channel 4, the signal is hardly detectable). The fit of the background data
for the three channels produces: $0.064C(t)=181.2-0.043t$ (channel 1); $%
0.064C(t)=138.6-0.039t$ (channel 2); $0.064C(t)=125.2-0.053t$ (channel 3).
With the method mentioned above, we find the primary value and position of
the peak count rate for the three channels being $176.2$ and $7.936s$
(channel 1), $261.5$ and $5.888s$ (channel 2), and $294.6$ and $4.416s$
(channel 3), respectively, and the position of the $FWHM$ in the decay phase
being $20.67s$ (channel 1), $17.73s$ (channel 2), and $13.12s$ (channel 3),
respectively (where, when searching the primary value and position of the
peak count rate, data of sample 4 of channel 1 are smoothed with the DB3
wavelet in the level of the fourth-class, instead of the third-class,
decomposition, due to the much scatter of data in this channel). The
normalized and re-scaled light curves of the burst in the three channels are
shown in Fig. 8. We find that the profiles of the light curves in the three
channels do not show an obvious different, indicating that the
self-similarity character holds for this source.

Thus, we consider here only the case of channel 3. Comparing the data of
sample 7 of channel 3 (see the pluses in the upper panel of Fig. 8) of this
burst with those curves in Fig. 3, we find that the profile of the light
curve arising from the local pulse of (83) with a sufficiently large local
width is the most likely one accounting for the pulse observed. The light
curve of this form of local pulses is therefore employed to perform a fit to
sample 6. The main formula employed for performing the fit is equation (21),
where for the rest frame radiation form we adopt the Band function (54). As
the influence of the rest frame radiation form to the profile of light
curves is insignificant (see Fig. 7) we adopt $\alpha _0=-1$, $\beta
_0=-2.25 $ and $\nu _{0,p}=1keV$ to perform the fit. Since the profile of
light curves is not sensitive to the Lorentz factor, as suggested in section
3, we adopt $\Gamma =10$. In the same way, for the local pulse of (83) we
take $\tau _{\theta ,\min }=0$. To meet the data observed, we assign $%
C_0=2\pi R_c^3I_0/hcD^2$ and $t=t_1\tau +t_0$, where $C_0$, $t_1$ and $t_0$
are free parameters which would be determined by fit. The fitting parameters
obtained with the least square method are listed in Table 2, where one finds
that the probability of rejecting the null hypothesis is $P\ll 0.001$. This
suggests that the profile of the light curve of GRB 930612 could indeed be
accounted for by a fireball emitting with an exponentially decaying local
pulse.

What could be determined from this analysis? As suggested in last section,
the width of the local pulse is not sensitive to the profile of light curves
when the former is large enough (see Figs. 4 and 5). For the fitting curve
of this burst, we obtain $FWHM1/FWHM2=0.507$. According to Fig. 4, this
value of $FWHM1/FWHM2$ is not sensitive to $\sigma $, and therefore the
value of $\sigma $ obtained by the fit above is not well determined.
However, in this situation, the type (suddenly dimming or gradually dimming)
of the corresponding local pulse is sensitive to the profile of light
curves, and hence the light curve of GRB 930612 arising from a gradually
dimming local pulse could be concluded, assuming that the source is
undergoing the fireball stage.

It is noticed that, before performing the fit, the signal data are smoothed.
Does the conclusion still hold if the data are not smoothed? To provide an
answer to this, we simply calculate the $\chi ^2$ of the fitting curve with
sample 4, and obtain $\chi ^2=1281$. Taking $1308$ as the number of degrees
of freedom we find the corresponding probability, of rejecting the null
hypothesis, as $P<0.001$, indicating that the conclusion holds in this
situation. But it shows that the goodness of fit owes much to the smooth of
data (note that the smooth of data itself does not guarantee the goodness of
fit without introducing a proper curve for the fit). The normalized and
re-scaled fitting curve as well as the signal data without being smoothed
are presented in Fig. 8 as well.

In the same way, count rates of channel 3 of GRB 910721 (\#563), GRB 920925
(\#1956), GRB 941026 (\#3257), GRB 951019 (\#3875) and GRB 951102B (\#3892)
are fitted, where we take the same values of $\alpha _0$, $\beta _0$, $\nu
_{0,p}$, $\Gamma $ and $\tau _{\theta ,\min }$ adopted above to perform the
fits. Local pulse (83) is taken to make a fit to count rates for GRB 910721,
while for GRB 941026 and GRB 951102B, local pulse (85) with $\mu =1$ is
adopted, and for GRB 920925 and GRB 951019, local pulse (86) with $\mu =1$
is assumed. In determining the primary value and position of the peak count
rate, sample 4 of GRB 941026 is smoothed with the DB3 wavelet in the level
of the fourth-class decomposition, instead of the third-class decomposition
adopted above, due to the much scatter of data (in this way, the position of
the $FWHM$ in the decay phase can be better determined). For GRB 951019 and
GRB 951102B, data of sample 4 are smoothed with the DB3 wavelet in the level
of the second-class decomposition since they are less scatter.

Free parameters obtained by the fits are listed in Table 2 as well. We find
for GRB 951102B that the probability, of rejecting the null hypothesis, is $%
P<0.001$, while for the other four bursts, the probability is $P\ll 0.001$.
It suggests that profiles of the light curve of these bursts could indeed be
accounted for by the Doppler effect of fireballs when appropriate local
pulses are assumed. Count rate light curves of sample 6 and the
corresponding fitting curves of these sources are presented in Fig. 9, where
all the curves are normalized and re-scaled based on the value and the
position of the peak count rate and the $FWHM$ position in the decay phase
of the corresponding fitting curves, calculated in the same way adopted in
Fig. 3. From these fitting curves we get $FWHM1/FWHM2=0.377$, $0.819$, $%
0.382 $, $0.416$ and $0.605$ for GRB 910721, GRB 920925, GRB 941026, GRB
951019 and GRB 951102B, respectively. According to Fig. 4, parameters $\tau
_{\theta ,0}$ and $\tau _{\theta ,\max }$ obtained above are well determined
for GRB 941026, GRB 951019 and GRB 951102B, while for GRB 910721 and GRB
920925, one can determine the ranges of $0.1\leq \sigma \leq 1$ and $1<\tau
_{\theta ,\max }$, respectively.

\section{Discussion and conclusions}

The analysis in this paper is under the assumption that the curvature effect
is important. Count rate formula used could not be applied to the cases in
which the fireball surface is not (globally or locally) spherically
symmetric, where the curvature effect is not at work. It should be pointed
out that, as already known, the profile of the pulses observed could be well
represented by various pulse functions and then the curvature effect is not
a unique mechanism to account for it. The analysis of the profile of pulses
alone is not sufficient to tell if the curvature effect is important. To
find an answer to this, other efforts should be made.

As shown above, the profile of pulses of fireball sources is not sensitive
to the rest frame radiation form, and based on this we are able to perform
fits to the light curves of several bursts under the assumption that they
are undergoing the fireball stage. However, as suggested by (21), count
rates of different energy channels could be described by a single formula.
Could one perform fits to the four channel light curves observed by BATSE
with (21)? The answer is yes if all physical parameters of a source are
known. We find that, to account for different channel light curves, the rest
frame radiation form plays an important role. As the corresponding rest
frame radiation parameters are not available for us, we could not perform a
further investigation on the fits to the sources discussed above.

Nevertheless, it would make sense if only showing how equation (21) is at
work when several channel light curves of a source are concerned. Here, let
us try to fit the four channel light curves of GRB 951019 when adopting
various sets of the rest frame radiation parameters. The method is the same
as that adopted in last section, except that we deal with four channels,
instead of one. When taking $\alpha _0=-1$, $\beta _0=-2.25$, $\Gamma =100$,
$\tau _{\theta ,\min }=0$, $\tau _{\theta ,\max }=0.518$ and $\mu =1$, we
obtain $\nu _{0,p}=1.06keV$, $C_0=4.91$, $t_1=28800$ and $t_0=-0.583$, which
leads to $\chi ^2=1186$ (with the number of degrees of freedom being $1044$
). The probability is $P=0.00191$. When adopting $\alpha _0=-0.5$, $\beta
_0=-3.5$, $\Gamma =100$, $\tau _{\theta ,\min }=0$, $\tau _{\theta ,\max
}=0.518$ and $\mu =1$, we get $\nu _{0,p}=0.907keV$, $C_0=11.1$, $t_1=27600$
and $t_0=-0.562$, which produces $\chi ^2=779.2$ (the number of degrees of
freedom is the same). The corresponding probability is $P\ll 0.001$. (When
allowing $\alpha _0$ and $\beta _0$ to be free, the fit will be slightly
improved.) Presented in Fig. 10 are the fitting curves of the second case,
together with the observed data of the four channel light curves of the
source. It shows that, different channel light curves of a burst could
indeed be accounted for by a single formula. Relations between them might
mainly be due to the Doppler effect of fireballs. The $FWHM$ of the fitting
curves of the second case are related with energy by a power law of $\log
(FWHM/s)=0.38-0.24\log (E/keV)$. The index, $-0.24$, is different from $-0.4$
which was obtained previously (see, e.g., Fenimore et al. 1995). Note that,
if $-0.24$ could be convinced (e.g., when the adopted rest frame radiation
indexes are true), it is from a single burst, but $-0.4$ arises from the sum
of the $FWHM$ of the individual sources of a sample and these widthes depend
on the distribution of the rest frame radiation parameters.

Although if the power law index of GRB 951019 is $-0.24$ is still an open
question, a power law relationship between the width and energy holds for
this burst would be true, which is obviously displayed in Fig. 10. Under the
theory of the Doppler effect of fireballs, this phenomenon is naturally
explained. While photons emitted from the small area of the fireball surface
with $\theta \sim 0$ would be observed in higher energy channels due to the
Doppler effect, those radiated from the larger area of the fireball surface
must be observed in lower energy channels, and the latter must last a much
longer time due to the geometric delay. This, we suspect, might become a
useful approach to check if the curvature effect is at work for any bursts
concerned.

As can be deduced from previous studies (see, e.g., Paper I), due to the
Doppler effect of fireballs, neglecting the area of $\theta >1/\Gamma $
would lead a light curve with a cutoff tail, or a turn over, in its decay
phase, which we call a cutoff tail problem. This feature would be obvious
when the local pulse is short enough, and under this circumstance, the
feature would become a criterion to pick out those sources emitted from the
area of $\theta <1/\Gamma $, from others (note that, as the count rate at $%
\theta =1/\Gamma $ is a quarter of the peak, the feature would be obviously
observable). When the local pulse is long, the cutoff tail, or the turn
over, feature would be less obvious and even be no more visible. For the
case of a patch moving along the direction of $\theta \sim 1/\Gamma $, the
light curve also exhibits the feature of cutoff tails when the local pulse
is short enough (see Fig. 1). Compared with that of the patch moving towards
the observer, its light curve lasts a much longer time, while the amplitude
becomes much smaller. When the local pulse lasts a sufficient interval of
time, the cutoff tail, or the turn over, feature would no longer be visible
for this patch, but instead, a full structure of FRED would be observed (see
Fig. 2). As noted by Ryde and Svensson (2002), there are some bursts that
their light curves have a sudden change, going into a more rapid decay. In
terms of the curvature effect, this turn over feature could be interpreted
as the light curves coming from the radiation of hot spots.

As shown in Fig. 4, when the local pulse width is small enough, the ratio of
the width of the light curve in the rising portion to that in the decay
phase would be sensitive to it, and in this situation, the ratio could be
well determined by fit. However, when the local pulse width is large, the
ratio would remains unchanged and the two quantities are no more uniquely
related and then this method would fail.

Replacing $\Gamma =10$ with $\Gamma =100$ when calculating some curves
discussed above, we find that the profile of light curves of fireballs is
not significantly affected by the Lorentz factor, suggesting that
conclusions referring to profiles would be maintained when different values
of $\Gamma $ within this range are considered. Thus, the character of FRED
as a consequence of the Doppler effect of fireballs is independent of the
Lorentz factor as long as the factor is large enough to represent a
relativistic motion. In addition, as shown in Figs. 3 and 6, the character
could result from any forms of local pulses and from any rest frame
radiation forms.

The interval $\Delta \tau _{pb}$ between the observed beginning and the peak
of the light curve of a local rectangle pulse is proportional to the local
width of the pulse (see Appendix B for a detailed analysis). For a large
value of the Lorentz factor, the peak count rate $C_p$ of the light curve of
local rectangle pulses would be proportional to $1/\Gamma ^4\beta ^2(1-\beta
)^3$. With the two quantities we get
\begin{equation}
\frac{C_p}{\Delta \tau _{pb}}\propto \frac{\Gamma ^4}{\Delta \tau _\theta }%
\quad \quad (\Gamma \gg 1).
\end{equation}
It indicates that the slope of the up rising part of an FRED pulse, if it
arises from a local pulse with a constant emission, would be very sensitive
to the Lorentz factor and be sensitive to the width of the local pulse as
well. Therefore, quantity $C_p/\Delta \tau _{pb}$ of pulses might be useful
for detecting the expanding speed of GRBs.

As is shown above, our analysis focuses on the model of fireballs which are
highly symmetric and expand relativistically. However, since the derivation
does not rely on any assumptions of the Lorentz factor, the basic formulas
(those in section 2) are applicable to sub-relativistic cases as well as
non-relativistic cases, as long as the objects concerned are highly
symmetric and are isotropically expanding. In our derivation, the thickness
of the outer shell is not taken into account. This does not matter. In the
analysis, the concept of the surface intensity is employed. Any radiation
from the shell must pass though the surface and at any time there is a
unique value of radiation passing through it, and this is the quantity
defined as the surface intensity. In this way, all radiations from or behind
the shell are included.

It should be noticed that the formula presented in this paper is applicable
to the radiation emitted from small areas such as $\theta \leq 1/\Gamma $ as
long as the areas concerned are locally highly symmetric. If all GRBs are
beamed, the discussion of the radiation emitted from the whole fireball
surface would become meaningless. However, since count rate light curves of
all the GRBs observed so far vary enormously, we suspect that there might be
various models accounting for all of these objects. Due to the great output
rate of the radiation observed, many GRBs would undergo the fireball stage
and some of them might probably be observed when they remain in this stage.
For a burst arising from the collapse of some massive objects, the
consequent fireball could become highly symmetric. The emitting area would
be the whole fireball surface when the radiation occurs before the fireball
shell is distorted, while it would be a patch (or a hot spot) when a short
inner jet hits the outer shell.

We suspect that, if during some period of time, continuous explosion inside
the fireball lasts an interval of time and its intensity keeps unchanged,
then the local pulse would be approximated by a rectangle one, as long as
the cooling time scale is short enough. Under the situation that the
radiation seeds (e.g., electrons) are distributed within the outer shell
geometrically with a Gaussian form and the inner shock occurred is quite
strong so that both inner and outer electrons gain the sam amount of energy
from the shock, a local Gaussian pulse might be produced (also, the cooling
time scale is assumed to be short enough).

In fitting the count rates of the six GRBs, we have very few free parameters
for each of them. It is plain that, when allowing other parameters such as
the indexes of the Band function to be variable, one would get much better
fits. However, in last section we focus on the question that if there are
any GRBs that the profiles of their count rate light curves can be described
by the count rate formula provided. If the formula can explain the observed
profiles when adopting some simple forms of the intensity of local pulses
and some certain values of the corresponding quantities (in fact, as the $%
\chi ^2$ shows, this is true), our task will be reached. As pointed out
above, the profile of count rate light curves of fireballs is not sensitive
to the Lorentz factor as long as the factor is large enough to represent a
relativistic motion. Thus, adopting $\Gamma =10$ is not fatal for the
goodness of fit of the six GRBs (one can check that adopting other values of
$\Gamma $ would also produce well fits for these sources). Note that, in
fitting the count rates of these GRBs, the cosmological effect is ignored
due to the lack of the knowledge of redshifts. While the change of the
magnitude of the light curves when taking into account the cosmological
effect can be absorbed into the magnitude itself, the frequency shifting of
the effect would affect the values of the quantities associated with $\tau $%
. However, the cosmological factor which is $1+z$ can be absorbed into these
quantities as well, and in this way the fitting curves in Figs. 8 and 9 will
not be affected.

As is mentioned above, the count rate formula presented in this paper is
derived in detail which does not rely on any approximately valued quantities
or estimated methods. Therefore it would be generally applicable. When some
factors are ignored, it will come to previous formulas such as those
presented in Papers I and II. A constraint of applying the formula is that
the object concerned must be one emitted within a locally highly symmetric
area which move outwards isotropically relative to the center of the object,
such as a cone expanding towards to the observer. Due to the vast difference
between various light curves of GRBs observed, we believe that the shape of
the light curves must vary significantly from source to source, and for many
GRBs the above constraint might not be satisfied. Therefore, to study
statistical properties of GRBs, it would be better to employ empirical or
semi-empirical functions such as those presented in Norris et al. (1996) and
Kocevski et al. (2003). Compared with those empirical functions, the formula
presented in this paper is more suitable to be applied to individual sources
when their count rate light curves are seen to be likely affected by the
curvature effect (e.g., if there exists a structure of FRED in all the well
separated pulses of the sources).

\newpage

\begin{center}
{\Large \textbf{Acknowledgments}}\\[0pt]
\end{center}

This work was supported by the Special Funds for Major State Basic Research
Projects (``973'') and National Natural Science Foundation of China (No.
10273019).

\vspace{20mm}


{\bf{\LARGE Appendix A. Relation between the peak count rate and
the width of $C_0(\tau )$ }}

Here, we employ the concept of $FWHM$ to describe the width of $C_0(\tau )$.

One can verify from (36) that the maximum value of $C_0(\tau )$ would be
obtained when $\tau \rightarrow (1-\beta )\tau _{\theta ,0}$. Let
\[
C_{0,p}\equiv C_0[\tau \rightarrow (1-\beta )\tau _{\theta ,0}].\qquad
\qquad \qquad \qquad \qquad \qquad \qquad \qquad \qquad \qquad \qquad \qquad
\qquad (A1)
\]
Then, from (36) we obtain
\[
C_{0,p}=\frac{1+\beta \tau _{\theta ,0}}{\Gamma ^3(1-\beta )^2}.\qquad
\qquad \qquad \qquad \qquad \qquad \qquad \qquad \qquad \qquad \qquad \qquad
\qquad \qquad \qquad (A2)
\]

Applying (A2), when $\beta >0$, one can obtain from (36) that
\[
\tau _H=\frac{-(1+k+\beta \tau _{\theta ,0})+(1+\beta \tau _{\theta ,0})%
\sqrt{k^2+(1+k)^2}}{k^2},\qquad \qquad \qquad \qquad \qquad \qquad \qquad
\qquad (A3)
\]
where
\[
C_0(\tau =\tau _H)=\frac{C_{0,p}}2.\qquad \qquad \qquad \qquad \qquad \qquad
\qquad \qquad \qquad \qquad \qquad \qquad \qquad \qquad \qquad (A4)
\]
Therefore, the width, described by the concept of $FWHM$, of the light curve
of the local $\delta $ function pulse would be
\[
\Delta \tau _{FWHM}=\tau _H-(1-\beta )\tau _{\theta ,0}=\frac{(\Gamma -\sqrt{%
\Gamma ^2-1})(\sqrt{2\Gamma ^2-1}-\Gamma )}{\Gamma ^2-1}(1+\beta \tau
_{\theta ,0}).\qquad \qquad \qquad (A5)
\]

From (39) we learn that the interval of the observable time of the local $%
\delta $ function pulse is
\[
\Delta \tau =1+\tau _{\theta ,0}-(1-\beta )\tau _{\theta ,0}=1+\beta \tau
_{\theta ,0}.\qquad \qquad \qquad \qquad \qquad \qquad \qquad \qquad \qquad
\qquad (A6)
\]
According to Paper III, the radius of the fireball at time $t_\theta $ can
be determined by
\[
R(t_\theta )=(t_\theta -t_c)\beta c+R_c.\qquad \qquad \qquad \qquad \qquad
\qquad \qquad \qquad \qquad \qquad \qquad \qquad \qquad \qquad (A7)
\]
Inserting (32) into (A7) we find
\[
1+\beta \tau _{\theta ,0}=\frac{R(t_{\theta ,0})}{R_c}.\qquad \qquad \qquad
\qquad \qquad \qquad \qquad \qquad \qquad \qquad \qquad \qquad \qquad \qquad
\qquad (A8)
\]
Thus, $\Delta \tau $ represents, in a relative term, the time scale of the
real size of the fireball at the corresponding emission time, $t_{\theta ,0}$%
. Applying (A8) we get from (A5) that
\[
\Delta \tau _{FWHM}=\frac{(\Gamma -\sqrt{\Gamma ^2-1})(\sqrt{2\Gamma ^2-1}%
-\Gamma )}{\Gamma ^2-1}\frac{R(t_{\theta ,0})}{R_c}.\qquad \qquad \qquad
\qquad \qquad \qquad \qquad \qquad (A9)
\]

From (A2) one finds
\[
C_{0,p}=\frac 1{\Gamma (\Gamma -\sqrt{\Gamma ^2-1})^2}\frac{R(t_{\theta ,0})%
}{R_c},\qquad \qquad \qquad \qquad \qquad \qquad \qquad \qquad \qquad \qquad
\qquad \qquad (A10)
\]
where, (A8) is applied. Combining (A9) and (A10) we get
\[
C_{0,p}=\frac{\Gamma ^2-1}{\Gamma (\Gamma -\sqrt{\Gamma ^2-1})^3(\sqrt{%
2\Gamma ^2-1}-\Gamma )}\Delta \tau _{FWHM}.\qquad \qquad \qquad \qquad
\qquad \qquad \qquad \qquad (A11)
\]
This is the relation between the peak count rate and the width of $C_0(\tau
) $.

Inserting (A2) into (36) yields
\[
\frac{C_0(\tau )}{C_{0,p}}=\frac{(\frac{\Delta \tau }\beta -\frac 1k-\tau
)\Delta \tau }{(1+k\tau )^2},\qquad \qquad \qquad \qquad \qquad \qquad
\qquad \qquad \qquad \qquad \qquad \qquad \qquad (A12)
\]
where (A6) is applied. With this, the ratio of a certain count rate to the
peak count rate of $C_0(\tau )$ for any observation time is determined. An
important application of this is to consider the case of $\theta _{\max
}=1/\Gamma $, for which we obtain the maximum value of $\tau $ from (51).
Applying this value to (A12) and assuming $\Gamma \gg 1$ we get
\[
C_0(\tau |_{\theta =1/\Gamma })\simeq \frac{C_{0,p}}4.\qquad \qquad \qquad
\qquad \qquad \qquad \qquad \qquad \qquad \qquad \qquad \qquad \qquad \qquad
\qquad (A13)
\]

\vspace{20mm}

{\bf{\LARGE Appendix B. Peak count rate of the light curve of
local rectangle pulses ignoring the rest frame spectral form}}

Here we present a detailed study on the peak count rate of the light curve
of a local rectangle pulse with its rest frame spectrum being a $\delta $
function form, for which the count rate is determined by (71).

We consider the case of the whole fireball surface for which (38) is
applied. In this case (72), (73) and (74) are applicable. From (74) we find
that, if
\[
(1-\beta )\tau _{\theta ,\max }<1+\tau _{\theta ,\min },\qquad \qquad \qquad
\qquad \qquad \qquad \qquad \qquad \qquad \qquad \qquad \qquad \qquad \qquad
(B1)
\]
there will be three ranges of $\tau $: $I_I\equiv \{(1-\beta )\tau _{\theta
,\min }\leq \tau \leq (1-\beta )\tau _{\theta ,\max }\}$, $I_{II}\equiv
\{(1-\beta )\tau _{\theta ,\max }\leq \tau \leq 1+\tau _{\theta ,\min }\}$,
and $I_{III}\equiv \{1+\tau _{\theta ,\min }\leq \tau \leq 1+\tau _{\theta
,\max }\}$. If
\[
1+\tau _{\theta ,\min }<(1-\beta )\tau _{\theta ,\max },\qquad \qquad \qquad
\qquad \qquad \qquad \qquad \qquad \qquad \qquad \qquad \qquad \qquad \qquad
(B2)
\]
there will be three other ranges of $\tau $: $II_I\equiv \{(1-\beta )\tau
_{\theta ,\min }\leq \tau \leq 1+\tau _{\theta ,\min }\}$, $II_{II}\equiv
\{1+\tau _{\theta ,\min }\leq \tau \leq (1-\beta )\tau _{\theta ,\max }\}$,
and $II_{III}\equiv \{(1-\beta )\tau _{\theta ,\max }\leq \tau \leq 1+\tau
_{\theta ,\max }\}$.

One can check that, in range $I_I$, $\widetilde{\tau }_{\theta ,\min }=\tau
_{\theta ,\min }$ and $\widetilde{\tau }_{\theta ,\max }=\tau /(1-\beta )$,
and then we get from (71) that
\[
\begin{array}{c}
C(\tau )= \\
\frac{2\pi R_c^3I_0}{5\nu _0hcD^2\Gamma ^4\beta ^2(1-\beta )^3}\{\frac{%
(5\beta -1)(1+k\tau )^2}4+\frac{5(1-\beta )(1+\beta \tau _{\theta ,\min })^4%
}{4(1+k\tau )^2}-\frac{(1+\beta \tau _{\theta ,\min })^5}{(1+k\tau )^3}\};
\end{array}
\qquad \qquad \qquad \qquad \qquad (B3)
\]
in range $I_{II}$, $\widetilde{\tau }_{\theta ,\min }=\tau _{\theta ,\min }$
and $\widetilde{\tau }_{\theta ,\max }=\tau _{\theta ,\max }$, and then we
get
\[
\begin{array}{c}
C(\tau )= \\
\frac{2\pi R_c^3I_0}{5\nu _0hcD^2\Gamma ^4\beta ^2(1-\beta )^3}\{\frac{%
(1+\beta \tau _{\theta ,\max })^5-(1+\beta \tau _{\theta ,\min })^5}{%
(1+k\tau )^3}-\frac{5[(1+\beta \tau _{\theta ,\max })^4-(1+\beta \tau
_{\theta ,\min })^4]}{4(1-\beta )^{-1}(1+k\tau )^2}\};
\end{array}
\qquad \qquad \qquad \qquad (B4)
\]
in range $I_{III}$, $\widetilde{\tau }_{\theta ,\min }=\tau -1$ and $%
\widetilde{\tau }_{\theta ,\max }=\tau _{\theta ,\max }$, and then we get
\[
\begin{array}{c}
C(\tau )= \\
\frac{2\pi R_c^3I_0}{5\nu _0hcD^2\Gamma ^4\beta ^2(1-\beta )^3}\{\frac{%
(1+\beta \tau _{\theta ,\max })^5}{(1+k\tau )^3}-\frac{5(1-\beta )(1+\beta
\tau _{\theta ,\max })^4}{4(1+k\tau )^2}+\frac{(1-\beta )^5(1+k\tau )^2}4\};
\end{array}
\qquad \qquad \qquad \qquad \qquad (B5)
\]
in range $II_I\equiv $, $\widetilde{\tau }_{\theta ,\min }=\tau _{\theta
,\min }$ and $\widetilde{\tau }_{\theta ,\max }=\tau /(1-\beta )$, then we
get (B3); in range $II_{II}$, $\widetilde{\tau }_{\theta ,\min }=\tau -1$
and $\widetilde{\tau }_{\theta ,\max }=\tau /(1-\beta )$, and then we get
\[
C(\tau )=\frac{2\pi R_c^3I_0}{5\nu _0hcD^2\Gamma ^4\beta ^2(1-\beta )^3}%
\frac{5\beta -1+(1-\beta )^5}4(1+k\tau )^2;\qquad \qquad \qquad \qquad
\qquad \qquad (B6)
\]
in range $II_{III}$, $\widetilde{\tau }_{\theta ,\min }=\tau -1$ and $%
\widetilde{\tau }_{\theta ,\max }=\tau _{\theta ,\max }$, and then we get
(B5).

Here, we pay our attention to relativistic motions, and hence we assume
\[
\beta >\frac 35\qquad \qquad \qquad \qquad \qquad \qquad \qquad \qquad
\qquad \qquad \qquad \qquad \qquad \qquad \qquad \qquad \qquad \qquad (B7)
\]
in the following analysis.

a) In the case of $(1-\beta )\tau _{\theta ,\max }<1+\tau _{\theta ,\min }$,
we have (B3) in range $I_I$, (B4) in range $I_{II}$ and (B5) in range $%
I_{III}$. Differentiating (B3), (B4), and (B5) we obtain
\[
\begin{array}{c}
\frac{dC(\tau )}{d\tau } \\
=\frac{\pi R_c^3I_0}{5\nu _0hcD^2\Gamma ^4\beta (1-\beta )^4}\frac{%
\{[(5\beta -1)(1+k\tau )^4-5(1-\beta )(1+\beta \tau _{\theta ,\min
})^4](1+k\tau )+6(1+\beta \tau _{\theta ,\min })^5\}}{(1+k\tau )^4},
\end{array}
\qquad \qquad \qquad (B8)
\]
\[
\begin{array}{c}
\frac d{d\tau }C(\tau ) \\
=\frac{\pi R_c^3I_0}{5\nu _0hcD^2\Gamma ^4\beta (1-\beta )^4}\frac{%
\{5[(1+\beta \tau _{\theta ,\max })^4-(1+\beta \tau _{\theta ,\min
})^4](1-\beta )(1+k\tau )-6[(1+\beta \tau _{\theta ,\max })^5-(1+\beta \tau
_{\theta ,\min })^5]\}}{(1+k\tau )^4},
\end{array}
\qquad (B9)
\]
and
\[
\begin{array}{c}
\frac d{d\tau }C(\tau ) \\
=\frac{\pi R_c^3I_0}{5\nu _0hcD^2\Gamma ^4\beta (1-\beta )^4}\frac{%
\{[(1-\beta )^4(1+k\tau )^4+5(1+\beta \tau _{\theta ,\max })^4](1-\beta
)(1+k\tau )-6(1+\beta \tau _{\theta ,\max })^5\}}{(1+k\tau )^4},
\end{array}
\qquad \qquad \qquad (B10)
\]
respectively.

One finds $\beta \tau _{\theta ,\min }\leq k\tau \leq \beta \tau _{\theta
,\max }$ in range $I_I$, $(1-\beta )(1+k\tau )\leq 1+\beta \tau _{\theta
,\min }$ in range $I_{II}$, and $(1-\beta )(1+k\tau )\leq (1+\beta \tau
_{\theta ,\max })$ in range $I_{III}$. Therefore,
\[
\begin{array}{c}
(5\beta -1)(1+k\tau )^4-5(1-\beta )(1+\beta \tau _{\theta ,\min })^4 \\
\geq 2(5\beta -3)(1+\beta \tau _{\theta ,\min })^4>0\quad \quad (\tau \in
I_I),
\end{array}
\qquad \qquad \qquad \qquad \qquad \qquad \qquad \qquad \qquad \qquad (B11)
\]
\[
\begin{array}{c}
5[(1+\beta \tau _{\theta ,\max })^4-(1+\beta \tau _{\theta ,\min
})^4](1-\beta )(1+k\tau )-6[(1+\beta \tau _{\theta ,\max })^5-(1+\beta \tau
_{\theta ,\min })^5] \\
\leq -[(1+\beta \tau _{\theta ,\max })^4-(1+\beta \tau _{\theta ,\min
})^4](1+\beta \tau _{\theta ,\min })\leq 0\quad \quad \quad \quad \quad
\quad \quad \quad (\tau \in I_{II}),
\end{array}
\quad (B12)
\]
\[
\lbrack (1-\beta )^4(1+k\tau )^4+5(1+\beta \tau _{\theta ,\max })^4](1-\beta
)(1+k\tau )-6(1+\beta \tau _{\theta ,\max })^5\leq 0\quad (\tau \in
I_{III}).\qquad (B13)
\]
In this case, $dC(\tau )/d\tau >0$ in range $I_I$, $dC(\tau )/d\tau \leq 0$
in range $I_{II}$, and $dC(\tau )/d\tau \leq 0$ in range $I_{III}$. Hence,
the peak of $C(\tau )$ must be located at the upper limit of $\tau $ in
range $I_I$. That is
\[
\begin{array}{c}
C_{I,p}=C[\tau =(1-\beta )\tau _{\theta ,\max }] \\
=\frac{2\pi R_c^3I_0}{5\nu _0hcD^2\Gamma ^4\beta ^2(1-\beta )^3}\{\frac{%
(1+\beta \tau _{\theta ,\max })^5-(1+\beta \tau _{\theta ,\min })^5}{%
(1+\beta \tau _{\theta ,\max })^3}-\frac{5(1-\beta )[(1+\beta \tau _{\theta
,\max })^4-(1+\beta \tau _{\theta ,\min })^4]}{4(1+\beta \tau _{\theta ,\max
})^2}\}.
\end{array}
\qquad \qquad (B14)
\]

b) In the case of $1+\tau _{\theta ,\min }<(1-\beta )\tau _{\theta ,\max }$,
we have (B3) in range $II_I$, (B6) in range $II_{II}$ and (B5) in range $%
II_{III}$. Differentiating (B3) and (B5) we obtain (B8) and (B10),
respectively. Differentiating (B6) we get
\[
\frac d{d\tau }C(\tau )=\frac{\pi R_c^3I_0}{5\nu _0hcD^2\Gamma ^4\beta
(1-\beta )^4}[(1-\beta )^5+5\beta -1](1+k\tau ).\qquad \qquad \qquad \qquad
\qquad \qquad (B15)
\]

We find $\beta \tau _{\theta ,\min }\leq k\tau $ in range $II_I$ and $%
(1-\beta )(1+k\tau )\leq (1+\beta \tau _{\theta ,\max })$ in range $II_{III}$%
. In the same way, one reaches $dC(\tau )/d\tau >0$ in range $II_I$ and $%
dC(\tau )/d\tau \leq 0$ in range $II_{III}$. Since $\beta >3/5$, we find
that
\[
(1-\beta )^5+5\beta -1>(1-\beta )^5+2>0.\qquad \qquad \qquad \qquad \qquad
\qquad \qquad \qquad \qquad \qquad \qquad \qquad (B16)
\]
Hence, $dC(\tau )/d\tau >0$ in range $II_{II}$. Therefore, the peak of $%
C(\tau )$ must be located at the upper limit of $\tau $ in range $II_{II}$.
That is
\[
\begin{array}{c}
C_{II,p}=C[\tau =(1-\beta )\tau _{\theta ,\max }] \\
=\frac{2\pi R_c^3I_0}{5\nu _0hcD^2\Gamma ^4\beta ^2(1-\beta )^3}\frac{%
[(1-\beta )^5+5\beta -1](1+\beta \tau _{\theta ,\max })^2}4.
\end{array}
\qquad \qquad \qquad \qquad \qquad \qquad \qquad \qquad \qquad
(B17)
\]

As shown above, when $\beta >3/5$, the position of the peak count rate of
the light curve of local rectangle pulses would be located at $\tau
=(1-\beta )\tau _{\theta ,\max }$. Therefore, according to (74), the
interval between the beginning and the peak of the pulse would be
\[
\Delta \tau _{pb}\equiv (1-\beta )\tau _{\theta ,\max }-(1-\beta
)\tau _{\theta ,\min }=(1-\beta )(\tau _{\theta ,\max }-\tau
_{\theta ,\min }).\qquad \qquad \qquad \qquad \qquad (B18)
\]
Let
\[
\Delta \tau _\theta \equiv \tau _{\theta ,\max }-\tau _{\theta ,\min
}.\qquad \qquad \qquad \qquad \qquad \qquad \qquad \qquad \qquad \qquad
\qquad \qquad \qquad \qquad \qquad (B19)
\]
One finds
\[
\Delta \tau _{pb}=(1-\beta )\Delta \tau _\theta .\qquad \qquad \qquad \qquad
\qquad \qquad \qquad \qquad \qquad \qquad \qquad \qquad \qquad \qquad \qquad
\qquad (B20)
\]
It suggests that the interval between the observed beginning and the peak of
the light curve of a local rectangle pulse is proportional to the local
width of the pulse. For a same kind of local rectangle pulses, $\Delta \tau
_{pb}$ would become an indicator of the Lorentz factor of the expanding
fireball.

From (B20) one finds that, when $\Delta \tau _\theta \rightarrow 0$, $\Delta
\tau _{pb}\rightarrow 0$. The profile would approach that of local $\delta $
function pulses.

From (B14) and (B17) we observe that, when $\Gamma \gg 1$, the peak of the
count rate would be proportional to $1/\Gamma ^4\beta ^2(1-\beta )^3$. Let $%
C_p$ be the observed peak count rate of a pulse. One can check that
\[
\frac{C_p}{\Delta \tau _{pb}}\propto \frac{\Gamma ^4}{\Delta \tau _\theta }
\quad \quad (\Gamma \gg 1).\qquad \qquad \qquad \qquad \qquad \qquad \qquad
\qquad \qquad \qquad \qquad \qquad \qquad \qquad (B21)
\]
This indicates that the slope of the up rising part of an FRED pulse, if it
can be described by the light curve of a local rectangle pulse, would be
very sensitive to the Lorentz factor and be sensitive to the width of the
local pulse as well.

Quantities $\Delta \tau _{pb}$ and $C_p/\Delta \tau _{pb}$ of
pulses might be useful for detecting the expanding speed of GRBs,
so long as the pulses can be described by the light curve of local
rectangle pulses.

\newpage

\begin{center}
\textbf{REFERENCES}
\end{center}

\begin{verse}
Band, D., et al. 1993, ApJ, 413, 281

Fenimore, E. E., Epstein, R. I., and Ho, C. 1993, A\&AS, 97,59

Fenimore, E. E., in't Zand, J. J. M., Norris, J. P., Bonnell, J. T., and
Nemiroff, R. J. 1995, ApJ, 448, L101

Fenimore, E. E., Madras, C. D., and Nayakshin, S. 1996, ApJ, 473, 998 (Paper
I)

Fishman, G. J., et al. 1994, ApJS, 92, 229

Ford, L. A., Band, D. L., Mattesou, J. L., et al. 1995, ApJ, 439, 307

Goodman, J. 1986, ApJ, 308, L47

Hailey, C. J., Harrison, F. A., and Mori, K. 1999, ApJ, 520, L25

Kocevski, D., Ryde, F., and Liang, E. 2003, ApJ, 596, 389

Krolik, J. H., and Pier, E. A. 1991, ApJ, 373, 277

M\'{e}sz\'{a}ros, P., and Rees, M. J. 1998, ApJ, 502, L105

Norris, J. P., Nemiroff, R. J., and Bonnell, J. T. et al. 1996, ApJ, 459, 393

Paczynski, B. 1986, ApJ, 308, L43

Preece, R. D., Pendleton, G. N., Briggs, M. S., et al. 1998, ApJ, 496, 849

Preece, R. D., Briggs, M. S., Mallozzi, R. S., et al. 2000, ApJS, 126, 19

Qin, Y.-P. 2002, A\&A, 396, 705 (Paper III)

Qin, Y.-P. 2003, A\&A, 407, 393

Ryde, F., and Petrosian, V. 2002, ApJ, 578, 290 (Paper II)

Ryde, F., and Svensson, R., ApJ, 566, 210

Schaefer, B. E., Teegaeden, B. J., Fantasia, S. F., et al. 1994, ApJS, 92,
285
\end{verse}

\newpage

\begin{center}
\textbf{Table 1: Typical values of $FWHM1/FWHM2$ withdrawn from Fig. 4}

\begin{tabular}{ccccccccc}
\hline\hline
$\sigma $ (or $\tau _{\theta, max} $, $\Delta \tau _{\theta, FWHM }$) & rect
& erd & er & ed & Gau & prd & pr & pd \\ \hline
0.01 & 0.02 & 0.11 & 0.03 & 0.12 & 0.07 & 0.06 & 0.03 & 0.09 \\
0.10 & 0.20 & 0.28 & 0.14 & 0.34 & 0.25 & 0.31 & 0.22 & 0.42 \\
1.00 & 1.07 & 0.35 & 0.24 & 0.44 & 0.37 & 0.58 & 0.57 & 0.79 \\
10.0 & 1.23 & 0.36 & 0.25 & 0.51 & 0.39 & 0.63 & 0.67 & 0.88 \\ \hline
\end{tabular}
\end{center}

Note: rect = rectangle; erd = exponential rise and decay; er = exponential
rise; ed = exponential decay; Gau = Gaussian; prd = power law rise and
decay; pr = power law rise; pd = power law decay

\vspace{30mm}

\begin{center}
\textbf{Table 2: Fitting parameters}

\begin{tabular}{ccccccc}
\hline\hline
Burst & $C_0$ & $t_1$ & $t_0$ & $\sigma$ & $\tau _{\theta,\max }(\tau
_{\theta ,0})$ & ($\chi^2$, n) \\ \hline
GRB 910721 & 1930 & 1200 & -0.658 & 0.197 &  & (325.7, 701) \\
GRB 920925 & 27.6 & 204 & -3.2 &  & 5.94 & (342.5, 434) \\
GRB 930612 & 248 & 242 & 0 & 2.11 &  & (785.4, 1308) \\
GRB 941026 & 2350 & 4110 & -1.34 &  & 0.244(0.0732) & (945.2, 1983) \\
GRB 951019 & 1010 & 218 & -0.489 &  & 0.518 & (145.4, 258) \\
GRB 951102B & 377 & 332 & -0.653 &  & 0.824(0.418) & (70.39, 117) \\ \hline
\end{tabular}
\end{center}

\end{document}